\documentstyle[12pt,aaspp4]{article}
\begin{document}

\title{Geodetic VLBI Observations of EGRET Blazars}

\author{B. Glenn Piner\altaffilmark{1,2}} 
\affil{NASA/Goddard Space Flight Center, Code 661, Greenbelt, MD 20771
}

\author{Kerry A. Kingham}
\affil{U.S. Naval Observatory, Earth Orientation Dept., 3450 Massachusetts Ave,
Washington D.C. 20392}

\altaffiltext{1}{Department of Astronomy, University of Maryland, 
College Park MD 20742}
\altaffiltext{2}{currently at: California Institute of Technology, Jet
Propulsion Laboratory, 4800 Oak Grove Dr., M-S: 238-332, Pasadena, CA
91109}

\begin{abstract}
We present VLBI observations of the EGRET quasars 0202+149, CTA 26, and 1606+106,
as well as additional analysis of VLBI observations of 1156+295 presented in Piner 
\& Kingham (1997b).  We have produced 8 and 2 GHz VLBI images at 11 epochs, 8 epochs,
and 12 epochs, spanning the years 1989 to 1996, of 0202+149, CTA 26, and 1606+106 respectively.    
The VLBI data have been taken from the Washington VLBI correlator's geodetic database.
We have measured the apparent velocities of the jet components and find that CTA 26 and
1606+106 are superluminal sources, with average apparent speeds of 8.9 and 2.9 $h^{-1}c$
respectively ($H_{0}=100h$ km s$^{-1}$ Mpc$^{-1}$, $q_{0}$=0.5).  The components in 0202+149
are stationary, and we identify this source as a compact F double.  These sources all
have apparently bent jets, and we detected non-radial motion of components in CTA 26 and 1156+295.
We have not yet detected any components emerging subsequent to the $\gamma$-ray flares in
CTA 26, 1156+295, and 1606+106, and we derive lower limits on the ejection times of any such components.
The misalignment angle distribution of the EGRET sources is compared to the distribution
for blazars as a whole, and we find that EGRET sources belong preferentially to neither the
aligned nor the misaligned population.  We also compare the average values for the apparent velocities
and Doppler beaming factors for the EGRET and non-EGRET blazars, and find no significant differences.
We thus find no indication, within the measurement errors, that EGRET blazars are any more 
strongly beamed than their counterparts which have not been detected in $\gamma$-rays. 
\end{abstract}

\section{Introduction}

Since its launch in 1991, the EGRET instrument on the 
{\em Compton Gamma Ray Observatory} (CGRO)
has detected with high significance 51 AGNs as emitters of 
high energy $\gamma$-rays (Mukherjee et al. 1997).  
These sources all appear to be members of the blazar class of AGNs 
(von Montigny et al. 1995a), containing BL Lac objects, highly
polarized quasars (HPQ), and optically violent variable (OVV) quasars.
In the radio these AGNs are radio-loud and flat-spectrum (radio spectral
index $\alpha \geq -0.5$).  The $\gamma$-ray emission from these
sources has notable properties, including high 
$\gamma$-ray flux that in many of the sources dominates
the flux at lower energies, and rapid variability on timescales
of a few days or less (e.g. von Montigny et al. 1995a).  The high $\gamma$-ray flux and rapid
time variability have been used to argue that the $\gamma$-ray emission
must be relativistically beamed; if not, the $\gamma$-ray emission
would be attenuated by pair-production optical depth.   Significant 
Doppler beaming factors for EGRET sources have been derived by this method
by e.g. Mattox et al. (1993) and Dondi \& Ghisellini (1995).

In the relativistic beaming model, the $\gamma$-ray sources are strongly
beamed and should display certain distinct properties
when imaged with VLBI, including
apparent superluminal motion of jet components, a high degree of core
dominance, and jets which are strongly bent by projection effects.
Some of the EGRET sources had been well monitored with VLBI before
their detection by EGRET, and they did indeed display these properties.
Proper motion measurements for 13 EGRET sources are
listed by Vermeulen \& Cohen (1994), hereafter VC94.
The measured apparent velocities of these sources are all superluminal
(provided the redshift of 0716+714 is above 0.28),
with the exceptions of 0458-020, 1127-145, and CTA 102.  VLBI observations
of other sources have been spurred by their EGRET detections, and superluminal
motion has recently been reported in 0420-014 (Wagner et al. 1995), 0528+134
(Krichbaum et al. 1995; Pohl et al. 1995), 0954+658 and 1219+285 (Gabuzda et al. 1994),
1633+382 (Barthel et al. 1995), and 1730-130 (Bower et al. 1997).
A campaign to monitor the southern EGRET sources with VLBI is underway, with first results
reported by Tingay et al. (1996).

The fact that many of the EGRET sources were relatively obscure until their
detection by EGRET indicates that, despite the recent concentration on VLBI observations
of EGRET sources, there remain many $\gamma$-ray sources which have not been
well studied with VLBI.  
To the best of our knowledge,
before this study there were 21 EGRET sources with 
published VLBI proper motion measurements
(many of which were not of high reliability (VC94)), leaving 30
EGRET sources which had not been studied with VLBI at multiple epochs.  Some of these sources
had been imaged or had had a VLBI intensity measurement made, but
proper motion studies had not been published. 
The study of which this paper is a part has attempted
to examine as many EGRET sources as possible which had not been previously well
observed with VLBI, using the geodetic VLBI database of the Washington VLBI
correlator located at the U.S. Naval Observatory (USNO). 
Reports have already been published of the measurement of superluminal motion in the
quasar 1611+343 (Piner \& Kingham 1997a, hereafter Paper I) 
and the measurement of a more standard
superluminal velocity for the quasar 1156+295 (Piner \& Kingham 1997b, hereafter Paper II),
which had previously had a much larger measured superluminal velocity than 
any other source (McHardy et al. 1993, 1990).
Detailed observations of three
more sources are presented in this paper.  

A study such as this has several potential benefits.
It is important to understand
why some of the sources sharing the common characteristics of EGRET blazars
have not been detected in $\gamma$-rays.
One possible reason is that the
$\gamma$-ray emission may be more narrowly beamed than the radio emission
(von Montigny et al. 1995b).  Other possible reasons include intrinsic
differences between the sources and long timescale variability of the
$\gamma$-ray emission.
VLBI measurements of such things as apparent superluminal motion
can provide information on important quantities such as the angle of the
jet to the line-of-sight, and can help to address this question.
High-resolution VLBI observations can also potentially discern any
effects of $\gamma$-ray flares on the jet structure.

This study also demonstrates the usefulness of archival geodetic VLBI data 
for astrophysics.  Several authors have used geodetic VLBI data for astrophysical
purposes.  All imageable sources in a single geodetic experiment
were imaged by Charlot (1990), and geodetic VLBI images have been presented
in papers on the individual sources 0528+134 (Krichbaum et al. 1995, Pohl et al. 1995),
OJ287 (Vicente, Charlot, \& Sol 1996), 4C 39.25 (Alberdi et al. 1993; Fey, Eubanks, \& Kingham 1997),
3C273 (Charlot 1993), 3C345 (Tang, R\"{o}nn\"{a}ng, \& B\aa\aa th 1989), and BL Lac
(Tateyama et al. 1998).
Schalinski et al. (1993) and Britzen et al. (1994) discussed
a large-scale project to image many geodetic VLBI sources over many experiments.
The study of which this paper is a part has a similar large-scale scope, with over
one hundred images of six EGRET sources having been studied in detail (Piner 1998).
Using archival geodetic VLBI data
has several advantages.  Many sources are observed very frequently, allowing for
excellent time sampling.  The archive extends back to 1986, allowing for
an approximate ten-year time baseline for most sources.
Many of these sources were not observed by the astronomical
VLBI community until after their announcement as EGRET sources in at least 1991.
We discuss the selection of the individual sources studied in this paper
in $\S$~\ref{sources}, 
the archived geodetic VLBI observations in $\S$~\ref{observations},
the motions of the individual jet components in $\S$~\ref{motion}, 
and astrophysical implications of these results in $\S$~\ref{discussion}.
We use $H_{0}=100h$ km s$^{-1}$ Mpc$^{-1}$, $q_{0}$=0.5, and
the flux $S\propto\nu^{+\alpha}$ throughout the paper.

\section {Selection of Individual Sources}
\label{sources}
For this study, we wished to find EGRET sources which had many good observations in the
USNO geodetic VLBI database and had not had VLBI proper motion measurements published.
For the list of EGRET sources, we used the 43 strong detections of AGNs given in the
second EGRET catalog (Thompson et al. 1995) and the catalog supplement (Thompson et al. 1996).
We also included four sources --- 3C66A, CTA 26, 1622-297, and 2155-304 ---
for which new results gave significant enough detections that
there was little doubt they would be included as strong detections in future lists
(Hartman 1996, private communication).
These four sources are indeed included, along with four other new sources,
in the recent list in Mukherjee et al. (1997).

Observations of 700 galactic and extragalactic sources are included
in the USNO geodetic VLBI database from the beginning of the program
in 1986 until 1996 November.
Of the 47 EGRET sources considered, 39 had at least one
observation in the database.  Of these, 26 had at least one observation with the
necessary $(u,v)$ plane coverage to make an image, and 20 of these had been observed
often enough and well enough to produce a series of images.  Nine of these 20 sources
already had reliable VLBI proper motion measurements published.  The remaining 11 sources were
0202+149, 0208-512, 0235+164, CTA 26, 0537-441, 1156+295, 1510-089, 1606+106, 1611+343, 
1622-253, and 1739+522.  Two of these sources, 0208-512 and 0537-441, are being studied
by Tingay et al. (1996), and we have already presented results on 1611+343 and 1156+295
in Papers I and II respectively.  Of the remaining seven sources, we present detailed
time series of 0202+149, CTA 26, and 1606+106 in this paper, and we have
also produced some images of 0235+164, 1622-253, and 1739+522.

\section{Observations}
\label{observations}
The VLBI observations used in this paper are from
archived geodetic Mark III VLBI observations processed at
the Washington VLBI Correlator Facility located at the U.S. Naval Observatory (USNO).
The background details of
this archival data and the reduction techniques are described in Paper I.
As discussed in Paper I, some images have been formed by combining data from several
geodetic VLBI experiments close together in time.  This can be done as long as there is
negligible change in the observable source structure over the time span of the summed
observations, which is true for all observations that were combined for this paper.
When experiments were combined, the AIPS task DBCON was used to combine them as separate subarrays.

A total of 40 different antennas were used among all of the imaged
observations presented in this paper, with a maximum of 11 being used for a 
single experiment.  The individual geodetic VLBI experiments used to image
0202+149, CTA 26, and 1606+106 are listed in Table~\ref{obslog}.  When two
or more experiments are listed at the same epoch, then these experiments were combined
to produce an image at this epoch.  When this was done
the time coordinate used for that epoch
was defined to be the average of the times of the
individual observations, weighted by the number of measured visibilities per observation.
We have observations over an approximate seven-year time baseline for 0202+149,
a three-year time baseline for CTA 26, and a five-year time baseline for 1606+106.
The source 0202+149 was imaged at 11 epochs at both 8 and 2 GHz, for a total of
22 images of this source.  Similarly, 16 and 24 images were produced of CTA 26 and 
1606+106 respectively.  A total of 60 geodetic VLBI experiments were analyzed to
produce the 62 images of these three sources.

\begin{table}
\caption{Observation Log of Geodetic VLBI Experiments Used}
\label{obslog}
{\scriptsize \begin{tabular}{c c c c c c c c c c} \tableline \tableline
\multicolumn{4}{c}{ } & \multicolumn{2}{c}{0202+149} &
\multicolumn{2}{c}{CTA 26} & \multicolumn{2}{c}{1606+106} \\
Exp\# \tablenotemark{a} & Name \tablenotemark{a} & Date & Antennas\tablenotemark{b} & 
Vis. \tablenotemark{c} & Epoch \tablenotemark{d} & Vis. & Epoch & Vis. & Epoch \\ \tableline
5970 & NAVY30  & 1989 Jul 23 & Gi,Mp,N8,Ri   &  900  & 1 & ...& ... & ... & ... \\
5971 & NAVY31  & 1989 Aug 4  & Gi,Ka,Mo,N8,Ri &  600  & 1 & ...& ... & ... & ... \\
5972 & NAVY32  & 1989 Aug 7  & Gi,Mp,N8,Ri & 1700  & 1 & ...& ... & ... & ... \\
5973 & NAVY33  & 1989 Aug 17 & Gi,Mp,N8,Ri & 2000  & 1 & ...& ... & ... & ... \\
5974 & NAVY34  & 1989 Aug 24 & Gi,Mp,N8,Ri & 1800  & 1 & ...& ... & ... & ... \\
6414 & NAVY114 & 1991 Mar 6  & Gi,Ka,N8,Ri &  700  & 2 & ...& ... & ... & ... \\
6415 & NAV115  & 1991 Mar 11 & Gi,Ka,N8,Ri & 1200  & 2 & ...& ... & ... & ... \\
6416 & NAV116  & 1991 Mar 20 & Gi,Ka,N8,Ri &  900  & 2 & ...& ... & ... & ... \\
6418 & NAV118  & 1991 Apr 3  & Gi,Ka,N8,Ri & 1200  & 2 & ...& ... & ... & ... \\
6419 & NAV119  & 1991 Apr 9  & Gi,Ka,N8,Ri &  700  & 2 & ...& ... & ... & ... \\  
6555 & GLOBAL   & 1991 Dec 17 & Gi,Ha,Ho,Ka,Km,Se,We,Wt          
& ... & ... & ... & ... & 6900  & 1  \\
7982 & POLAR-N1 & 1992 Jun 29 & D1,Gi,Hy,Ks,Me,On                
& ... & ... & ... & ... & 900   & 2  \\
6554 & PPM-S1   & 1992 Jul 15 & D4,Gi,Ka,Ks,Sa,Se                
& ... & ... & ... & ... & 2700  & 2  \\
7992 & GLOBAL   & 1992 Jul 21 & Gi,Ha,Ho,Ka,Ks,Sa,Se,We,Wt       
& ... & ... & ... & ... & 12300 & 2  \\
5789 & SN1BS9   & 1993 Apr 21 & Br,Gi,Ka,La & ... & ... & 2800 & 1 & ... & ...  \\
6191 & NAEX31   & 1993 May 17 & Gi,Ha,Ka,Ks,Ma,N8,Sa          
& ... & ... & 2500 & 1 & ... & ...  \\
6214 & NAEX32   & 1993 Jun 14 & Al,Gi,Ha,Ho,Ka,Ko,Ks,Ma,N8,Sa 
& 2400  & 3 & 1900 & 1 & ... & ...  \\
6573 & NEOSB002 & 1993 Jun 24 & Al,Gi,Ma,N8,Wt & ... & ... & ... & ... & 800   & 3  \\
6522 & SATL-ALT & 1993 Jun 25 & D6,Ha,Hn,Sa,Sc & ... & ... & ... & ... & 1100  & 3  \\
6187 & XASIA-2  & 1993 Jul 16 & D4,Ha,Ho,Ks,Se,Wt  
& ... & ... & ... & ... & 2700  & 3  \\
8069 & NAEX37   & 1993 Dec 6  & Gi,Ha,Ho,Ko,Ks,Ma,N8,Wt 
& ... & ... & 500  & 2 & ... & ...  \\
8024 & NEOSB008 & 1993 Dec 9  & Al,Ma,N8,On,Wt & ... & ... & 900  & 2 & ... & ... \\
6545 & NAXG07  & 1994 Jan 10 & Fo,Gi,Ko,Ma,N8,Wt             &  900  & 4 
& ... & ... & ... & ... \\
6591 & NAXG08  & 1994 Jan 12 & Al,Fo,Ha,Ho,Ka,Ks,Ma,N8       & 2300  & 4 
& ... & ... & ... & ... \\
6838 & NEOSA038 & 1994 Jan 18 & Br,Fo,Gi,Ko,Mk,N8,Sc,Wt          
& ... & ... & 1800 & 2 & 1500  & 4  \\
6905 & PPMS2    & 1994 Mar 16 & D4,Gi,Km,Ko,Se,Ur                
& ... & ... & ... & ... & 1200  & 5  \\
6931 & NEOSB011 & 1994 Mar 24 & Gi,Ma,N8,On,Wt                   
& ... & ... & ... & ... & 2100  & 5  \\
6933 & NEOSB013 & 1994 May 12 & Al,Gi,Ma,N8,On,Wt                
& ... & ... & ... & ... & 2100  & 5  \\
6936 & NEOSB016 & 1994 Sep 1  & Al,Gi,Ma,N8,On,Wt,Yl             
& ... & ... & 1900 & 3 & 2400  & 6  \\
6937 & NEOSB017 & 1994 Sep 8  & Al,Gi,Ma,N8,On,Wt,Yl             
& ... & ... & 1700 & 3 & 2300  & 6  \\
6875 & NEOSA075 & 1994 Oct 4  & Fo,Gi,Ko,N8,Ny,Wt                
& ... & ... & ... & ... & 2200  & 6  \\
6938 & NEOSB018 & 1994 Oct 5  & Al,Gi,Ma,N8,On,Wt,Yl             
& ... & ... & 2200 & 3 & 2800  & 6  \\
6939 & NEOSB019 & 1994 Oct 6  & Al,Gi,Ma,N8,On,Wt,Yl             
& ... & ... & 1100 & 3 & 3700  & 6  \\
6885 & NEOSA085 & 1994 Dec 13 & Fo,Gi,Ko,N8,Ny,Wt                
& ... & ... & ... & ... & 2400  & 7  \\
6020 & RDGTR1  & 1995 Jan 23 & Cr,Fo,Ha,Ho,Ma,Mk,N8,Pt,Sa,Sc 
& 4100  & 5 & ... & ... & ... &... \\
6891 & NA091    & 1995 Jan 24 & Fo,Gi,Ko,N8,Wt                   
& ... & ... & ... & ... & 1700  & 7  \\
6534 & RDNAP2   & 1995 Feb 1  & Al,Br,Nl,Ov,We                
& ... & ... & 800  & 4 & ... & ... \\
6541 & GTRF2    & 1995 Mar 6  & Cr,D4,Gi,Km,Ko,Ny,We,Wt          
& ... & ... & ... & ... & 2300  & 8  \\
6629 & PPMS1    & 1995 Mar 23 & D4,Gi,Ho,Km,Ko,Se             
& ... & ... &  900  & 5 & ... & ...  \\
6633 & RDTPC1   & 1995 Apr 20 & Br,Kp,Ks,Mk,Ov,Se,Ur          
& ... & ... &  700  & 5 & ... & ... \\
6909 & NAEXS6  & 1995 Jun 14 & Al,Gi,Mi,N8,Ny,Wt             
&  500  & 6 & ... & ... & ... & ... \\
6636 & RDGTR3  & 1995 Jul 24 & D6,Gi,Ha,Ho,Ko,Kp,Ks,Nt,Sa,Yl 
& 3800  & 7 & ... & ... & ... & ... \\
6780 & GTRF4    & 1995 Aug 16 & D6,Fo,Gi,Ha,Ho,Ko,Me,On,Sa,We,Yl 
& ... & ... & ... & ... & 4200  & 9  \\
6792 & RDC95A   & 1995 Aug 23 & Gi,Ko,Ny,On,We,Wt                
& ... & ... & ... & ... & 1200  & 9  \\
6990 & RDC95B   & 1995 Aug 24 & Gi,Ko,Ny,On,We,Wt                
& ... & ... & ... & ... & 1300  & 9  \\
6991 & RDC95C   & 1995 Aug 25 & Gi,Ko,Ny,On,We,Wt                
& ... & ... & ... & ... & 800   & 9  \\
6992 & RDC95D   & 1995 Aug 26 & Gi,Ko,Ny,On,We,Wt                
& ... & ... & ... & ... & 1300  & 9  \\
7080 & NAEX46   & 1995 Oct 11 & Al,Gi,Mi,N2,N8,We,Yl          
& ... & ... & 1600 & 6 & ... & ...  \\
8129 & NA129    & 1995 Oct 17 & Fo,Gi,Ko,Mk,N2,Nl,Wt          
& 4300 & 8 & 1300 & 6 & 1900 & 10 \\
8132 & NA132    & 1995 Nov 7  & Fo,Ko,Mi,N2,Wt                
& ... & ... & 1100 & 6 & ... & ...  \\
7311 & RDSATL   & 1996 Mar 11 & D6,Fo,Ha,Hn,Sc                
& ... & ... & 600  & 7 & ... & ...  \\
7312 & RDTPC    & 1996 Mar 11 & Br,Gi,Ks,Mk,Ov                
& ... & ... & 700  & 7 & 1800 & 11  \\
8152 & NA152   & 1996 Mar 26 & Fo,Gi,Ko,N2,Ny,Wt             
&  900  & 9 & ... & ... & ... & ... \\
\end{tabular}}
\end{table}

\begin{table}[!t]
Table 1 (cont.): Observation Log of Geodetic VLBI Experiments Used \\
{\scriptsize
\begin{tabular}{c c c c c c c c c c} \tableline \tableline
\multicolumn{4}{c}{ } & \multicolumn{2}{c}{0202+149} &
\multicolumn{2}{c}{CTA 26} & \multicolumn{2}{c}{1606+106} \\
Exp\# \tablenotemark{a} & Name \tablenotemark{a} & Date & Antennas\tablenotemark{b} & 
Vis. \tablenotemark{c} & Epoch \tablenotemark{d} & Vis. & Epoch & Vis. & Epoch \\ \tableline
7313 & WPAC     & 1996 Mar 27 & D4,Gi,Ko,Ks,Se                   
& ... & ... & ... & ... & 1100  & 11 \\
7316 & GTRF10   & 1996 Apr 23 & Cr,D1,Gi,Ha,Ho,Ks,Ma,Ny,We       
& ... & ... & ... & ... & 3100  & 11 \\
7321 & NAXS10  & 1996 Apr 24 & Al,Gi,Ko,N2,Ny,Wt             
& 9500  & 10 & ... & ... & ... & ... \\
7323 & RDNAP1   & 1996 Jul 11 & Al,Gi,Nl,We,Yl                   
& ... & ... & 1000 & 8 & 2300  & 12 \\
7317 & RDGT11   & 1996 Jul 22 & Br,D6,Fo,Ha,Ho,Ks,Mk,Ny,Sc       
& ... & ... & ... & ... & 2100  & 12 \\
7328 & RDNAP2   & 1996 Jul 22 & Al,Gi,Nl,We,Yl                   
& ... & ... & 800 & 8 & 2300  & 12 \\
8175 & NA175   & 1996 Sep 3  & Fo,Gi,Ko,N2,Ny,Wt             
& 1300  & 11 & ... & ... & ... & ... \\ \tableline
\end{tabular}}
\tablenotetext{a}{{\scriptsize Experiment number and name as it appears in experiment summary file.}}
\tablenotetext{b}{{\scriptsize Antenna names, locations, and sizes are as follows:
Al = ALGOPARK; Algonquin, Ontario; 46 m --- Br = BR-VLBA; Brewster, WA; 25 m --- 
Cr = CRIMEA; Simeiz, Crimea, Ukraine; 22 m --- D1 = DSS15; Goldstone, CA; 34 m ---
D4 = DSS45; Tidbinbilla, Australia; 34 m --- D6 = DSS65; Madrid, Spain; 34 m ---
Fo = FORTLEZA; Fortaleza, Brazil; 14 m --- Gi = GILCREEK; Fairbanks, AK; 26 m ---
Ha = HARTRAO; Hartebeesthoek, South Africa; 26 m --- Hn = HN-VLBA; Hancock, NH; 25 m ---
Ho = HOBART26; Hobart, Tasmania; 26 m --- Hy = HAYSTACK; Westford, MA; 37 m ---
Ka = KAUAI; Kauai, HI; 9 m --- Km = KASHIM34, KASH\_34; Kashima, Japan; 34 m ---
Ko = KOKEE; Kauai, HI; 20 m --- Kp = KP-VLBA; Kitt Peak, AZ; 25 m ---
Ks = KASHIMA; Kashima, Japan; 26 m --- La = LA-VLBA; Los Alamos, NM; 25 m ---
Ma = MATERA; Matera, Italy; 20 m --- Me = MEDICINA; Medicina, Italy; 32 m ---
Mi = MIAMI20; Perrine, FL; 20 m --- Mk = MK-VLBA; Mauna Kea, HI; 25 m ---
Mo = MOJAVE12; Goldstone, CA; 12 m --- Mp = MARPOINT; Maryland Point, MD; 26 m ---
N2 = NRAO20; Greenbank, WV; 20 m --- N8 = NRAO85\_3; Greenbank, WV; 26 m ---
Nl = NL-VLBA; North Liberty, IA; 25 m --- Nt = NOTO; Noto, Sicily, Italy; 25 m ---
Ny = NYALES20; Ny Alesund, Norway; 20 m --- On = ONSALA60; Onsala, Sweden; 20 m ---
Ov = OV-VLBA; Big Pine, CA; 25 m --- Pt = PIETOWN; Pietown, NM; 25 m ---
Ri = RICHMOND; Perrine, FL; 18 m --- Sa = SANTIA12; Santiago, Chile; 12 m ---
Sc = SC-VLBA; St. Croix, US Virgin Islands; 25 m --- Se = SESHAN25; Shanghai, China; 25 m ---
Ur = URUMQI; Xinjiang, China; 25 m --- We = WESTFORD; Westford, MA; 18 m ---
Wt = WETTZELL; Wettzell, Germany; 20 m --- Yl = YLOW7296; Yellowknife, Northwest Territory; 10 m.}}
\tablenotetext{c}{{\scriptsize Number of measured visibilities per IF, rounded to the nearest hundred.  
To get total number of visibilities
multiply by 8 for 8 GHz or 6 for 2 GHz.}}
\tablenotetext{d}{{\scriptsize Duplicate epoch numbers indicate combined experiments.}}
\end{table}

In addition to  the geodetic VLBI images, we have also used some VLBA images made as
part of a study of the Radio Reference Frame (Johnston et al. 1995).  
These images are also at 8 and
2 GHz, with some additional images at 15 GHz.  The images are courtesy of
Fey (1996, 1997, private communications), but we performed the model fitting of the images
independently.  VLBA observations of other Radio Reference Frame sources
are presented by Fey, Clegg, \& Fomalont (1996) and Fey \& Charlot (1997).  A list
of the VLBA observations used in this paper is given in Table~\ref{vlba}.  In general, the VLBA
images have lower resolution due to their shorter baselines, 
but higher dynamic range than the geodetic VLBI images.
Thus they can serve as useful consistency checks on the reality of fainter components
but do not significantly improve the proper motion measurements.

\begin{table}[!h]
\caption{VLBA Radio Reference Frame Observations Used}
\label{vlba}
\begin{tabular}{c c c c} \tableline \tableline
& & Frequencies & \\
Date & Sources & (GHz) & Epoch \tablenotemark{a} \\ \tableline
1995 Jul 24 & CTA 26, 1606+106 & 2,8 & F1 \\
1995 Oct  2 & CTA 26 & 2,8 & F2 \\
1995 Oct 17 & 0202+149, 1606+106 & 15 & F3 \\
1996 Apr 23 & 0202+149, 1606+106 & 2,8,15 & F4 \\ \tableline
\end{tabular}
\tablenotetext{a}{Epoch identification for comparison with geodetic VLBI epochs.}
\end{table}

\subsection{Images}
Figures 1 to 6 show time-series mosaics of some of the geodetic VLBI images, at 
2 and 8 GHz, of 0202+149, CTA 26, and 1606+106.  These figures display time along the
$y$-axis and relative right ascension along the $x$-axis.  The images are 
centered at the time coordinate at which the observation was made.  The axis scalings
have been chosen so that the images remain true to the original beam, and no distortion
is introduced into the images.  Because of space considerations, only about half of the VLBI
images used in the analysis of these sources are shown in these figures. 
We believe we have selected those images which best demonstrate
changes in source structure.
Parameters of the displayed images are given in Table~\ref{imtab}.  
The lowest contour for each image has been set equal to three times the rms noise
in the image; however, noise contours have been
suppressed in the time-series plots to avoid plotting a noise contour over an
adjacent image.  

\begin{figure}
\plotfiddle{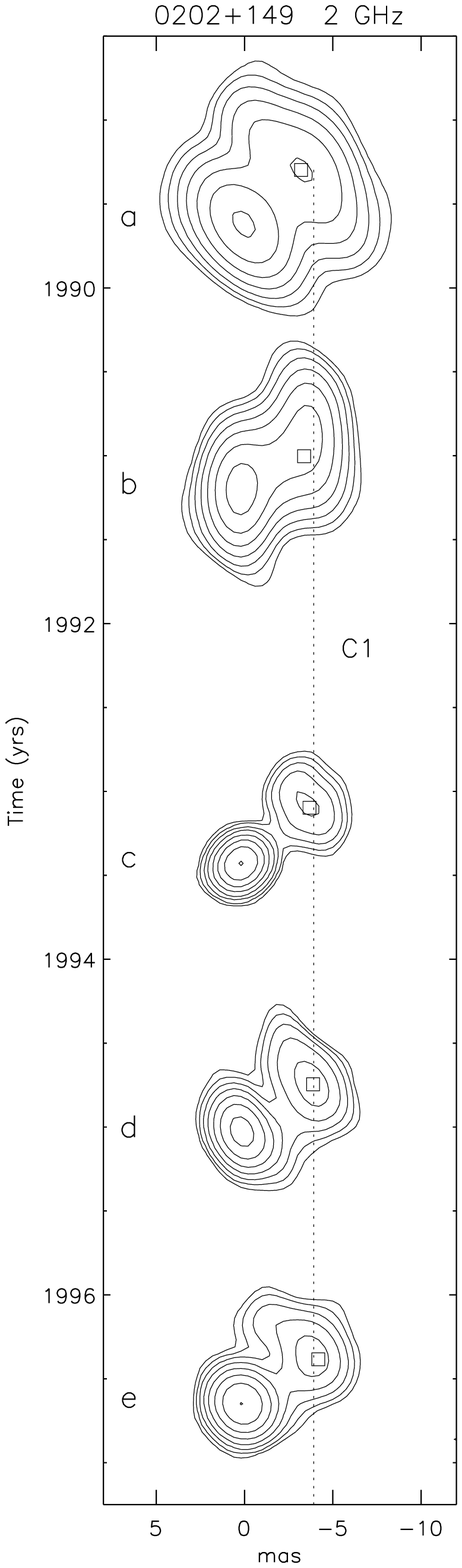}{7.125 in}{0}{73}{73}{-327}{-12}
\plotfiddle{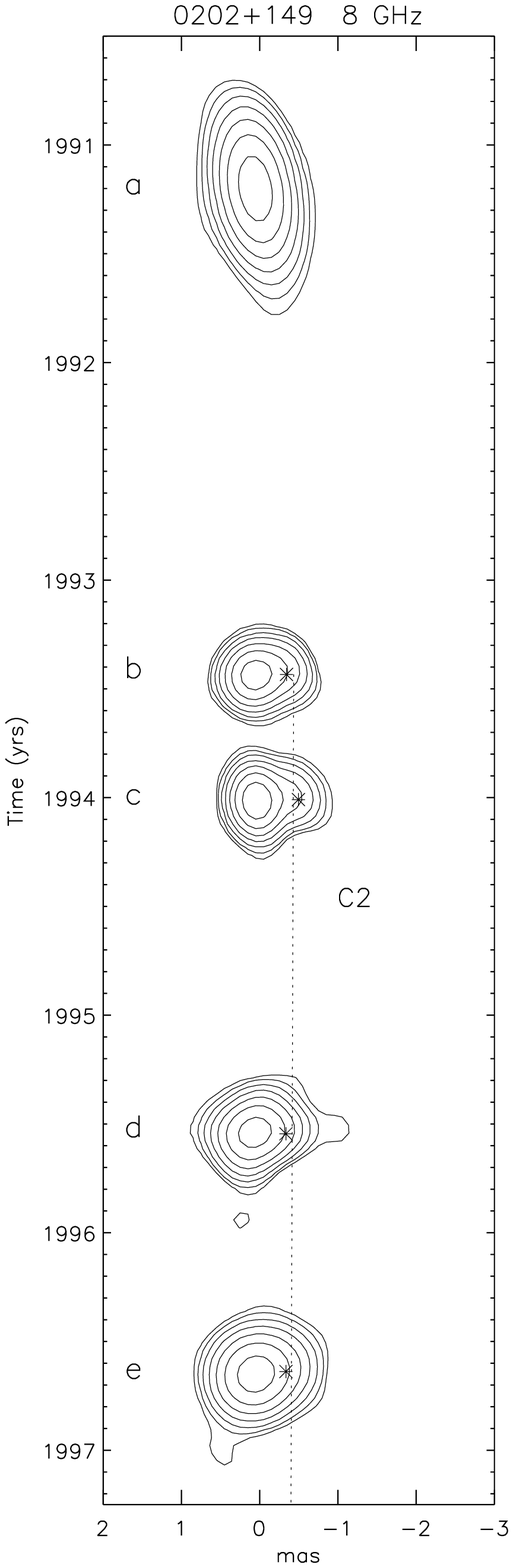}{0.0 in}{0}{73}{73}{90}{18}
Figures 1 and 2: Time-series mosaics of 0202+149 images at 2 and 8 GHz
respectively.
\end{figure}
\begin{figure}
\plotfiddle{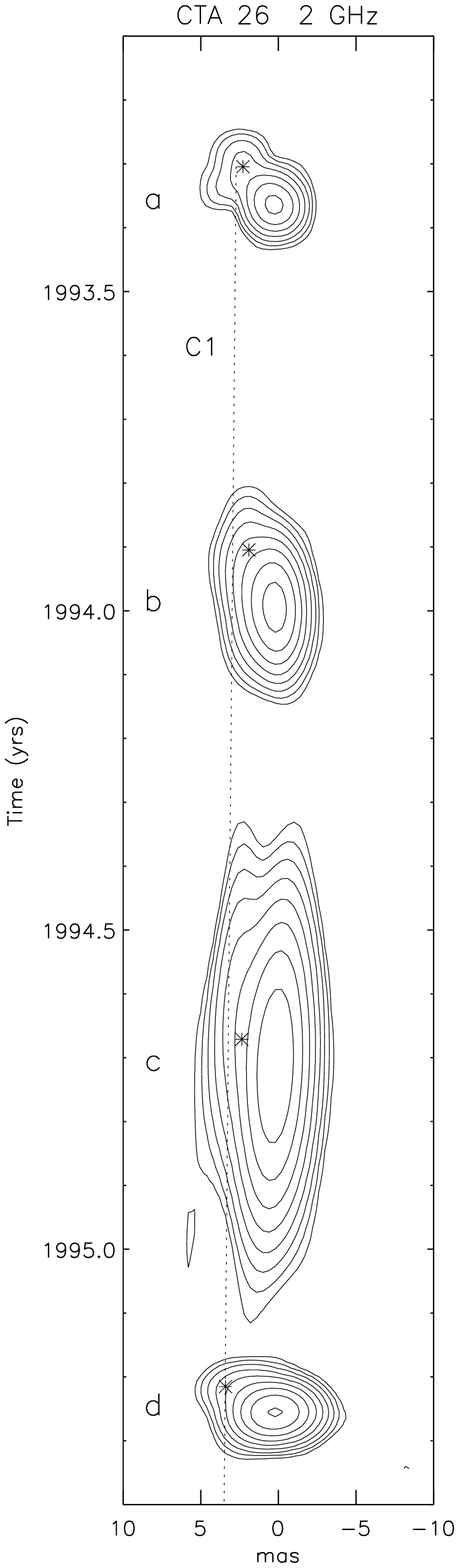}{7.125 in}{0}{73}{73}{-327}{-12}
\plotfiddle{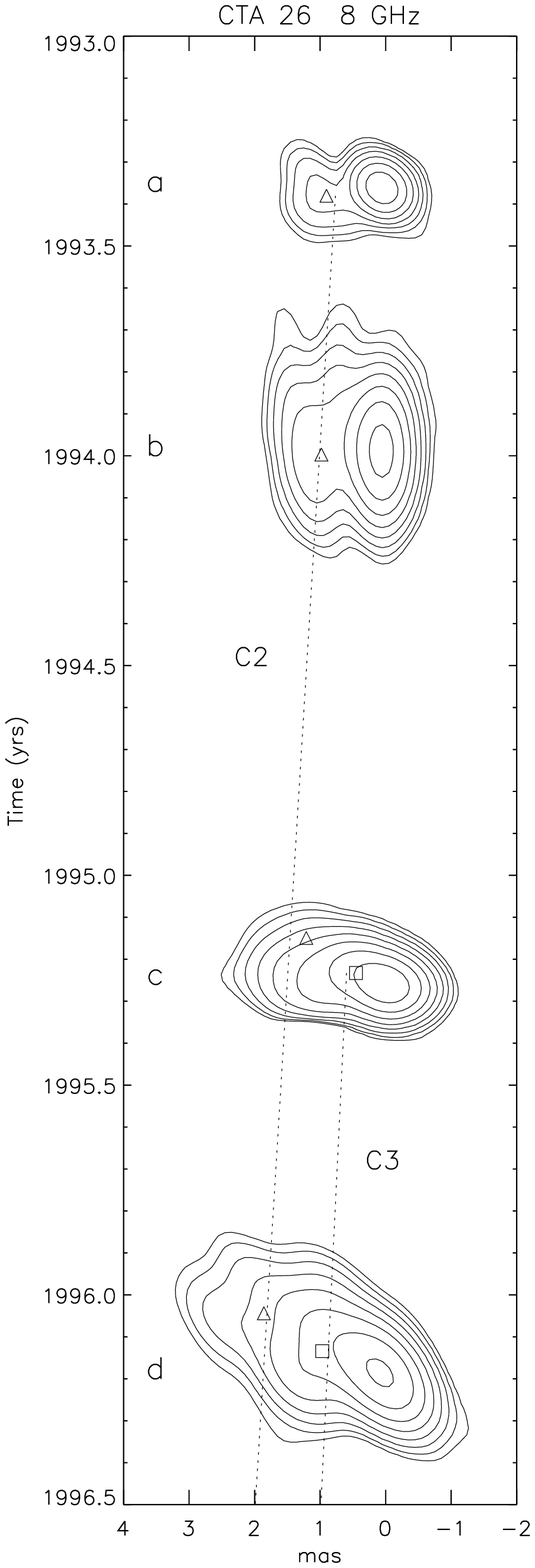}{0.0 in}{0}{73}{73}{90}{18}
Figures 3 and 4: Time-series mosaics of CTA 26 images at 2 and 8 GHz
respectively.
\end{figure}
\begin{figure}
\plotfiddle{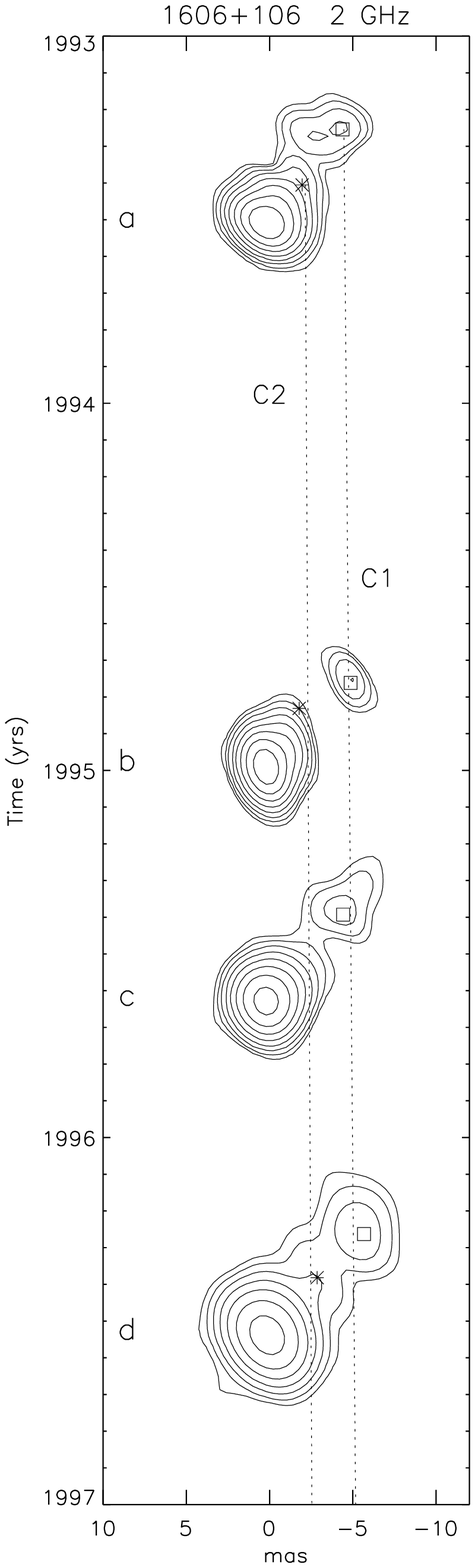}{7.125 in}{0}{73}{73}{-327}{-12}
\plotfiddle{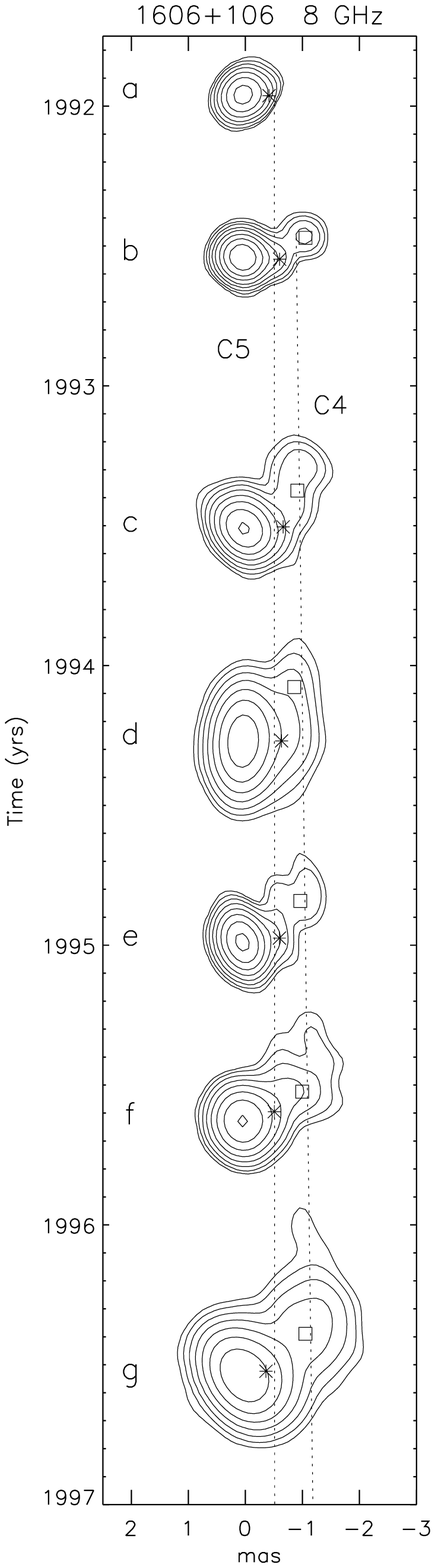}{0.0 in}{0}{73}{73}{90}{18}
Figures 5 and 6: Time-series mosaics of 1606+106 images at 2 and 8 GHz
respectively.
\end{figure}
\begin{table}
\caption{Parameters of the Displayed Images}
\label{imtab}
\begin{tabular}{l c c c l l c} \tableline \tableline
& & & & & Contours & Lowest \\
& Frequency & & & & (multiples of & Contour\tablenotemark{c} \\
Source & (GHz) & Figure & Epoch\tablenotemark{a} & Beam\tablenotemark{b} &
lowest contour) & (mJy beam$^{-1}$) \\ \tableline
0202+149 & 2 & 1$a$ & 1 & 4.56,2.70,31.8 & 1,2,4,8,16,32,64 & 17.0 \\
& & 1$b$ & 2 & 3.89,2.23,3.2 & 1,2,4,8,16,32,64 & 11.5 \\
& & 1$c$ & 3 & 1.92,1.68,-43.8 & 1,2,4,8,16,32,64,128 & 11.1 \\
& & 1$d$ & 5 & 2.59,1.76,25.0 & 1,2,4,8,16,32,64 & 9.5 \\
& & 1$e$ & 11 & 2.33,2.12,29.7 & 1,2,4,8,16,32,64,128 & 11.0 \\
& 8 & 2$a$ & 2 & 1.00,0.54,5.9 & 1,2,4,8,16,32,64 & 23.0 \\
& & 2$b$ & 3 & 0.48,0.43,1.6 & 1,2,4,8,16,32,64 & 16.8 \\
& & 2$c$ & 4 & 0.52,0.39,6.7 & 1,2,4,8,16,32,64 & 17.7 \\
& & 2$d$ & 7 & 0.53,0.45,-60.5 & 1,2,4,8,16,32,64 & 16.1 \\
& & 2$e$ & 11 & 0.61,0.57,14.9 & 1,2,4,8,16,32,64 & 15.6 \\
CTA 26 & 2 & 3$a$ & 1 & 2.07,1.84,26.3 & 1,2,4,8,16,32,64 & 26.6 \\
& & 3$b$ & 2 & 4.64,1.84,2.5 & 1,2,4,8,16,32,64,128 & 12.1 \\
& & 3$c$ & 3 & 10.6,2.11,-2.8 & 1,2,4,8,16,32,64,128 & 8.7 \\
& & 3$d$ & 5 & 2.69,1.95,-80.2 & 1,2,4,8,16,32,64,128,256 & 9.4 \\
& 8 & 4$a$ & 1 & 0.56,0.48,32.1 & 1,2,4,8,16,32,64 & 23.9 \\
& & 4$b$ & 2 & 1.24,0.50,-0.5 & 1,2,4,8,16,32,64,128 & 9.0 \\
& & 4$c$ & 5 & 0.70,0.53,-79.0 & 1,2,4,8,16,32,64,128 & 4.2 \\
& & 4$d$ & 7 & 1.05,0.66,35.6 & 1,2,4,8,16,32,64,128 & 6.4 \\
1606+106 & 2 & 5$a$ & 3 & 2.37,1.64,46.6 & 1,2,4,8,16,32,64,128 & 7.5 \\
& & 5$b$ & 7 & 2.50,1.53,17.0 & 1,2,4,8,16,32,64,128 & 7.7 \\
& & 5$c$ & 9 & 2.53,2.00,22.9 & 1,2,4,8,16,32,64,128 & 8.5 \\
& & 5$d$ & 12 & 3.52,2.44,36.5 & 1,2,4,8,16,32,64 & 18.0 \\
& 8 & 6$a$ & 1 & 0.51,0.44,-31.6 & 1,2,4,8,16,32,64 & 12.3 \\
& & 6$b$ & 2 & 0.48,0.45,55.8 & 1,2,4,8,16,32,64,128 & 7.1 \\
& & 6$c$ & 3 & 0.64,0.45,39.8 & 1,2,4,8,16,32,64,128 & 7.8 \\
& & 6$d$ & 5 & 1.07,0.59,-2.0 & 1,2,4,8,16,32,64 & 9.2 \\
& & 6$e$ & 7 & 0.65,0.42,18.8 & 1,2,4,8,16,32,64,128 & 7.5 \\
& & 6$f$ & 9 & 0.69,0.54,15.9 & 1,2,4,8,16,32,64,128 & 8.0 \\
& & 6$g$ & 12 & 1.02,0.68,33.2 & 1,2,4,8,16,32,64 & 9.8 \\ \tableline
\end{tabular}
\tablenotetext{a}{Epoch identification in Table~\ref{obslog}.}
\tablenotetext{b}{Numbers given for the beam are the FWHMs of the major
and minor axes in mas, and the position angle of the major axis in degrees.}
\tablenotetext{c}{The lowest contour is set to be three times the rms noise
in the image.}
\end{table}

The individual component positions as determined from model fitting
(see $\S$~\ref{mfittext}) are plotted on Figures 1 to 6, using various symbols.  The dotted
lines connecting these component positions represent the best fits to motion of
the components at constant velocity, using measured positions from 
{\em all} epochs. 
Proper motion of components is easy to see by looking at these fits: A perfectly
vertical line indicates a stationary component, and the more the line deviates
from the vertical the faster the component is moving.
Throughout the rest of this paper we follow the component numbering system of labeling the
presumed core C0 and labeling the other components consecutively starting at
C1, from the outermost component inward.  Many of the differences in
appearance from image to image within a series are due to the differences in
the geodetic VLBI experiments.  Experiments can vary drastically in both the
lengths of the baselines and the sensitivities of the antennas.  The geodetic VLBI 
experiments tend to have very long baselines which give high resolution; however,
some experiments also have a dearth of short baselines, which
can lead to an insensitivity to components which are farther out and more extended.

\subsection{Model Fits}
\label{mfittext}
In order to quantify the positions and motions of the jet components, we fit Gaussian
models to the observed visibilities for each epoch,  
using the MODELFIT procedure of the Caltech DIFMAP software package.
The fitted components were either
elliptical Gaussians, circular Gaussians, or delta functions.
Elliptical Gaussians were fit where possible; if the ellipses became very elongated
(axial ratio less than 0.1), 
then two or more circular Gaussians were used instead.  If the circular components
became very small (less than a tenth of a beam),  then a delta function was used.
Table~\ref{mfittab} lists the fluxes and positions of the major components
of the best fitting model for each image.  We have included model fits for all of the
observations listed in Table~\ref{obslog},
as well as for the VLBA images
listed in Table~\ref{vlba}.

\begin{table}
{\scriptsize 
\caption{Gaussian Models}
\label{mfittab}
\begin{tabular}{c c c c c c c c c c c c} \tableline \tableline
& & & Frequency & $S$\tablenotemark{b} & $r$\tablenotemark{c} & 
$\sigma_{r}$\tablenotemark{d} & PA\tablenotemark{c} &
$\sigma_{{\rm PA}}$\tablenotemark{d} & $a$\tablenotemark{e} & 
& $\Phi$\tablenotemark{f} \\
Source & Epoch\tablenotemark{a} & Component & (GHz) & (Jy) & (mas) & 
(mas) & (deg) & (deg) & (mas) & $b/a$ & (deg) \\ \tableline
0202+149 & 1 & C0 & 8 & 1.37 & ... & ... & ... & ... & 0.15 & 1.00 & ... \\
& & & 2 & 1.10 & ... & ... & ... & ... & ... & ... & ... \\
& & C1 & 2 & 0.98 & 4.58 & 0.28 & -44.8 & 9.3 & 4.26 & 0.42 & 44.6 \\
& 2 & C0 & 8 & 2.52 & ... & ... & ... & ... & 0.51 & 0.19 & 19.8 \\
& & C1 & 2 & 0.58 & 4.72 & 0.28 & -54.9 & 7.1 & 1.18 & 1.00 & ... \\
& 3 & C0 & 8 & 1.64 & ... & ... & ... & ... & 0.16 & 0.13 & -53.0 \\
& & C2 & 8 & 0.46 & 0.38 & 0.07 & -80.6 & 11.9 & ... & ... & ... \\
& & C1 & 2 & 0.59 & 4.97 & 0.19 & -47.9 & 3.2 & 2.53 & 0.80 & 44.5 \\
& 4 & C0 & 8 & 2.07 & ... & ... & ... & ... & 0.12 & 0.58 & -36.7 \\
& & C2 & 8 & 0.36 & 0.49 & 0.06 & -82.4 & 10.0 & ... & ... & ... \\
& & C1 & 2 & 0.40 & 4.90 & 0.17 & -57.4 & 3.3 & 1.94 & 0.16 & -26.3 \\
& 5 & C0 & 8 & 2.04 & ... & ... & ... & ... & 0.29 & 0.56 & -25.9 \\
& & C2 & 8 & 0.40 & 0.49 & 0.08 & -65.9 & 12.6 & 0.41 & 1.00 & ... \\
& & C1 & 2 & 0.62 & 4.78 & 0.18 & -53.2 & 5.1 & 2.98 & 0.45 & 39.4 \\
& 6 & C0 & 8 & 1.85 & ... & ... & ... & ... & 0.45 & 0.59 & -1.1 \\
& & C2 & 8 & 0.31 & 0.58 & 0.17 & -58.6 & 17.0 & ... & ... & ... \\
& & C1 & 2 & 0.77 & 4.31 & 0.40 & -51.8 & 7.3 & 3.73 & 0.68 & -22.1 \\
& 7 & C0 & 8 & 1.79 & ... & ... & ... & ... & 0.31 & 0.62 & -11.9 \\
& & C2 & 8 & 0.21 & 0.37 & 0.09 & -84.9 & 11.8 & ... & ... & ... \\
& & C1 & 2 & 0.34 & 4.89 & 0.21 & -60.2 & 3.0 & 1.55 & 0.83 & -48.5 \\
& 8 & C0 & 8 & 2.07 & ... & ... & ... & ... & 0.24 & 0.53 & -62.8 \\
& & C2 & 8 & 0.15 & 0.49 & 0.08 & -80.5 & 13.9 & ... & ... & ... \\
& & C1 & 8 & 0.23 & 4.82 & 0.14 & -48.5 & 2.0 & 2.48 & 0.22 & 55.0 \\
& & & 2 & 0.53 & 4.91 & 0.21 & -53.0 & 5.3 & 2.74 & 0.54 & 22.3 \\
& F3 & C0 & 15 & 1.87 & ... & ... & ... & ... & 0.34 & 0.17 & -19.7 \\
& & C2 & 15 & 0.26 & 0.34 & 0.12 & -79.6 & 29.5 & 0.39 & 1.00& ... \\
& & C1 & 15 & 0.21 & 4.99 & 0.17 & -53.8 & 3.4 & 2.42 & 0.52 & 40.4 \\
& 9 & C0 & 8 & 1.44 & ... & ... & ... & ... & ... & ... & ... \\
& & C2 & 8 & 0.37 & 0.34 & 0.08 & -63.6 & 20.0 & 0.42 & 1.00 & ... \\
& & C1 & 2 & 0.37 & 4.53 & 0.16 & -56.5 & 5.5 & 2.42 & 0.58 & 30.5 \\
& F4 & C0 & 15 & 1.52 & ... & ... & ... & ... & 0.27 & 0.23 & -26.9 \\ 
& & & 8 & 1.66 & ... & ... & ... & ... & ... & ... & ... \\
& & C2 & 15 & 0.28 & 0.39 & 0.15 & -50.9 & 21.1 & 0.64 & 0.29 & 76.8 \\
& & & 8 & 0.21 & 0.50 & 0.28 & -65.9 & 32.7 & 0.24 & 1.00 & ... \\
& & C1 & 15 & 0.21 & 5.12 & 0.22 & -54.0 & 2.6 & 2.41 & 0.41 & 36.2 \\
& & & 8 & 0.25 & 5.14 & 0.48 & -53.8 & 4.6 & 2.10 & 0.48 & 30.3 \\
& & & 2 & 0.41 & 4.49 & 0.67 & -57.4 & 13.1 & 2.55 & 0.53 & -2.1 \\
& 10 & C0 & 8 & 1.41 & ... & ... & ... & ... & 0.24 & 0.53 & -34.6 \\
& & C2 & 8 & 0.13 & 0.54 & 0.15 & -68.4 & 18.7 & ... & ... & ... \\
& & C1 & 8 & 0.12 & 5.11 & 0.28 & -52.5 & 2.5 & 1.76 & 0.38 & 20.0 \\
& & & 2 & 0.24 & 4.88 & 0.40 & -55.0 & 7.1 & 1.11 & 1.00 & ... \\
& 11 & C0 & 8 & 1.50 & ... & ... & ... & ... & 0.28 & 0.25 & -52.7 \\
& & C2 & 8 & 0.30 & 0.38 & 0.09 & -76.6 & 15.1 & 0.27 & 1.00 & ... \\
& & C1 & 2 & 0.34 & 4.93 & 0.21 & -56.7 & 4.5 & 2.10 & 0.37 & 17.9 \\
CTA 26 & 1 & C0 & 8 & 2.89 & ... & ... & ... & ... & 0.26 & 0.40 & 59.8 \\
& & C2 & 8 & 0.78 & 0.90 & 0.13 & 94.4 & 8.6 & 0.85 & 0.60 & 45.4 \\
& & C1 & 2 & 0.31 & 3.67 & 0.51 & 45.3 & 7.2 & ... & ... & ... \\
& 2 & C0 & 8 & 1.81 & ... & ... & ... & ... & 0.34 & 0.54 & 63.4 \\
& & C2 & 8 & 0.56 & 0.97 & 0.13 & 90.0 & 17.8 & 0.88 & 0.79 & -3.9 \\
& & C1 & 2 & 0.16 & 4.41 & 1.07 & 27.7 & 8.3 & 1.04 & 1.00 & ... \\
& 3 & C0 & 8 & 1.16 & ... & ... & ... & ... & ... & ... & ... \\
& & C3 & 8 & 0.66 & 0.48 & 0.04 & 90.3 & 20.8 & ... & ... & ... \\
& & C2 & 8 & 0.28 & 1.22 & 0.29 & 108.2 & 29.1 & ... & ... & ... \\
\end{tabular}}
\end{table}

\begin{table}
Table 4 (cont.): Gaussian Models \\
{\scriptsize
\begin{tabular}{c c c c c c c c c c c c} \tableline \tableline
& & & Frequency & $S$\tablenotemark{b} & $r$\tablenotemark{c} & 
$\sigma_{r}$\tablenotemark{d} & PA\tablenotemark{c} &
$\sigma_{{\rm PA}}$\tablenotemark{d} & $a$\tablenotemark{e} & 
& $\Phi$\tablenotemark{f} \\
Source & Epoch\tablenotemark{a} & Component & (GHz) & (Jy) & (mas) & 
(mas) & (deg) & (deg) & (mas) & $b/a$ & (deg) \\ \tableline
& & C1 & 2 & 0.26 & 3.65 & 1.36 & 58.3 & 32.6 & ... & ... & ... \\
& 4 & C0 & 8 & 0.99 & ... & ... & ... & ... & ... & ... & ... \\
& & C3 & 8 & 0.79 & 0.69 & 0.08 & 92.1 & 27.1 & ... & ... & ... \\
& & C2 & 8 & 0.22 & 1.74 & 0.42 & 76.1 & 38.7 & ... & ... & ... \\
& & C1 & 2 & 0.19 & 4.70 & 4.03 & 36.5 & 35.1 & 5.01 & 1.00 & ... \\
& 5 & C0 & 8 & 0.94 & ... & ... & ... & ... & 0.21 & 1.00 & ... \\
& & C3 & 8 & 0.64 & 0.62 & 0.04 & 64.7 & 3.4 & 0.42 & 1.00 & ... \\
& & C2 & 8 & 0.17 & 1.54 & 0.16 & 58.9 & 5.7 & 0.28 & 1.00 & ... \\
& & C1 & 2 & 0.08 & 4.27 & 0.61 & 63.7 & 7.5 & ... & ... & ... \\
& F1 & C0 & 8 & 0.93 & ... & ... & ... & ... & ... & ... & ... \\
& & C3 & 8 & 0.70 & 0.67 & 0.08 & 61.2 & 10.1 & 0.85 & 0.20 & -82.7 \\
& & C2 & 8 & 0.23 & 1.68 & 0.30 & 69.7 & 16.5 & 1.83 & 0.44 & 22.2 \\
& & C1 & 2 & 0.07 & 6.12 & 1.23 & 61.4 & 15.6 & 3.79 & 1.00 & ... \\
& F2 & C0 & 8 & 1.18 & ... & ... & ... & ... & ... & ... & ... \\
& & C3 & 8 & 0.63 & 0.79 & 0.08 & 66.5 & 9.6 & 0.92 & 0.60 & -67.5 \\
& & C2 & 8 & 0.18 & 1.64 & 0.35 & 58.2 & 17.5 & 1.89 & 0.63 & 13.4 \\
& 6 & C0 & 8 & 1.21 & ... & ... & ... & ... & 0.27 & 1.00 & ... \\
& & C3 & 8 & 0.63 & 0.81 & 0.05 & 62.8 & 7.6 & 0.88 & 0.36 & -85.7 \\
& & C2 & 8 & 0.13 & 1.95 & 0.21 & 58.5 & 12.4 & 0.68 & 1.00 & ... \\
& 7 & C0 & 8 & 1.07 & ... & ... & ... & ... & 0.52 & 0.51 & 51.0 \\
& & C3 & 8 & 0.53 & 1.03 & 0.06 & 67.9 & 2.8 & 1.02 & 0.52 & -23.9 \\
& & C2 & 8 & 0.14 & 2.08 & 0.25 & 62.5 & 5.2 & 0.96 & 1.00 & ... \\
& 8 & C0 & 8 & 0.76 & ... & ... & ... & ... & ... & ... & ... \\
& & C3 & 8 & 0.82 & 0.97 & 0.21 & 73.8 & 16.3 & 2.09 & 0.33 & ... \\
1606+106 & 1 & C0 & 8 & 1.16 & ... & ... & ... & ... & 0.14 & 1.00 & ... \\
& & C5 & 8 & 0.07 & 0.41 & 0.12 & -85.6 & 16.8 & ... & ... & ... \\
& 2 & C0 & 8 & 1.85 & ... & ... & ... & ... & 0.15 & 0.90 & 82.7 \\
& & C5 & 8 & 0.08 & 0.60 & 0.12 & -88.3 & 10.8 & ... & ... & ... \\
& & C4 & 8 & 0.06 & 1.12 & 0.11 & -69.7 & 5.9 & ... & ... & ... \\
& 3 & C0 & 8 & 1.37 & ... & ... & ... & ... & 0.31 & 0.68 & -47.6 \\
& & C5 & 8 & 0.08 & 0.64 & 0.13 & -83.6 & 13.1 & ... & ... & ... \\
& & C4 & 8 & 0.07 & 1.12 & 0.11 & -51.4 & 8.1 & 0.30 & 1.00 & ... \\
& & C3 & 8 & 0.04 & 1.65 & 0.12 & -36.6 & 5.5 & ... & ... & ... \\
& & C2 & 2 & 0.16 & 3.22 & 0.41 & -38.5 & 10.4 & 1.34 & 1.00 & ... \\
& & C1 & 2 & 0.08 & 7.32 & 0.41 & -37.2 & 4.6 & 1.31 & 1.00 & ... \\
& 4 & C0 & 8 & 0.91 & ... & ... & ... & ... & 0.18 & 0.60 & -7.1 \\
& & C5 & 8 & 0.13 & 0.51 & 0.15 & -58.8 & 27.5 & ... & ... & ... \\
& 5 & C0 & 8 & 1.01 & ... & ... & ... & ... & 0.27 & 0.73 & -40.3 \\
& & C5 & 8 & 0.20 & 0.64 & 0.15 & -84.6 & 22.6 & 0.63 & 1.00 & ... \\
& & C4 & 8 & 0.03 & 1.31 & 0.23 & -40.8 & 8.8 & ... & ... & ... \\
& & C2 & 8 & 0.05 & 4.08 & 0.22 & -45.7 & 3.0 & 0.44 & 1.00 & ... \\
& 6 & C0 & 8 & 1.16 & ... & ... & ... & ... & 0.15 & 0.27 & -61.0 \\
& & C5 & 8 & 0.15 & 0.44 & 0.11 & -79.0 & 19.8 & ... & ... & ... \\
& & C4 & 8 & 0.06 & 1.10 & 0.12 & -62.4 & 7.9 & 0.24 & 1.00 & ... \\
& & C2 & 8 & 0.07 & 4.31 & 0.30 & -36.0 & 3.3 & 0.39 & 1.00 & ... \\
& 7 & C0 & 8 & 1.32 & ... & ... & ... & ... & 0.24 & 0.54 & -81.6 \\
& & C5 & 8 & 0.07 & 0.61 & 0.11 & -78.4 & 14.9 & ... & ... & ... \\
& & C4 & 8 & 0.10 & 1.23 & 0.11 & -51.4 & 7.3 & 0.65 & 1.00 & ... \\
& & C3 & 8 & 0.02 & 1.91 & 0.14 & -26.1 & 4.1 & ... & ... & ... \\
& & C2 & 2 & 0.02 & 4.14 & 0.53 & -26.3 & 7.0 & ... & ... & ... \\
& & C1 & 2 & 0.06 & 7.18 & 0.45 & -43.2 & 4.6 & ... & ... & ... \\
& 8 & C0 & 8 & 1.53 & ... & ... & ... & ... & 0.26 & 0.46 & -46.8 \\
& & C5 & 8 & 0.19 & 0.33 & 0.12 & -83.6 & 23.1 & ... & ... & ... \\
\end{tabular}}
\end{table}

\begin{table}
Table 4 (cont.): Gaussian Models \\
{\scriptsize
\begin{tabular}{c c c c c c c c c c c c} \tableline \tableline
& & & Frequency & $S$\tablenotemark{b} & $r$\tablenotemark{c} & 
$\sigma_{r}$\tablenotemark{d} & PA\tablenotemark{c} &
$\sigma_{{\rm PA}}$\tablenotemark{d} & $a$\tablenotemark{e} & 
& $\Phi$\tablenotemark{f} \\
Source & Epoch\tablenotemark{a} & Component & (GHz) & (Jy) & (mas) & 
(mas) & (deg) & (deg) & (mas) & $b/a$ & (deg) \\ \tableline
& & C4 & 8 & 0.15 & 1.46 & 0.11 & -50.2 & 5.8 & 0.39 & 1.00 & ... \\
& & C1 & 2 & 0.05 & 8.01 & 0.43 & -37.4 & 3.7 & 1.46 & 1.00 & ... \\
& F1 & C3 & 8 & 0.03 & 2.35 & 0.44 & -29.9 & 7.9 & 1.01 & 1.00 & ... \\
& & C1 & 8 & 0.01 & 8.05 & 0.41 & -39.1 & 2.6 & 1.68 & 0.96 & 62.6 \\
& & & 2 & 0.07 & 8.44 & 1.54 & -35.5 & 8.6 & 4.52 & 0.33 & -33.8 \\
& 9 & C0 & 8 & 1.26 & ... & ... & ... & ... & 0.32 & 0.30 & -69.8 \\
& & C5 & 8 & 0.11 & 0.55 & 0.14 & -66.9 & 17.4 & 0.11 & 1.00 & ... \\
& & C4 & 8 & 0.10 & 1.14 & 0.14 & -60.5 & 8.5 & 0.65 & 0.10 & 59.5 \\
& & C3 & 8 & 0.05 & 1.97 & 0.15 & -36.1 & 4.6 & 0.88 & 1.00 & ... \\
& & C1 & 2 & 0.07 & 7.02 & 0.53 & -39.7 & 4.9 & 1.10 & 1.00 & ... \\
& 10 & C0 & 8 & 0.74 & ... & ... & ... & ... & 0.22 & 1.00 & ... \\
& & C5 & 8 & 0.12 & 0.58 & 0.21 & -70.1 & 23.3 & ... & ... & ... \\
& & C4 & 8 & 0.06 & 1.34 & 0.27 & -45.6 & 8.1 & 0.22 & 1.00 & ... \\
& & C1 & 2 & 0.08 & 7.55 & 1.22 & -33.7 & 4.9 & 4.25 & 0.25 & -87.8 \\
& F3 & C0 & 15 & 0.84 & ... & ... & ... & ... & 0.43 & 0.65 & -68.4 \\
& & C5 & 15 & 0.11 & 0.59 & 0.23 & -48.6 & 31.2 & 0.96 & 0.41 & 32.7 \\
& & C4 & 15 & 0.06 & 1.51 & 0.23 & -46.0 & 13.1 & 0.57 & 1.00 & ... \\
& 11 & C0 & 8 & 1.23 & ... & ... & ... & ... & 0.31 & 0.37 & -67.3 \\
& & C5 & 8 & 0.21 & 0.60 & 0.12 & -61.9 & 14.6 & 0.40 & 0.18 & 27.9 \\
& & C4 & 8 & 0.12 & 1.48 & 0.12 & -58.7 & 6.0 & ... & ... & ... \\
& & C3 & 8 & 0.04 & 1.84 & 0.12 & -42.2 & 4.8 & ... & ... & ... \\
& & C2 & 2 & 0.05 & 3.05 & 0.49 & -40.5 & 11.4 & ... & ... & ... \\
& & C1 & 2 & 0.04 & 8.27 & 0.49 & -44.2 & 4.2 & ... & ... & ... \\
& F4 & C0 & 15 & 0.71 & ... & ... & ... & ... & 0.21 & 0.68 & -63.7 \\
& & C5 & 15 & 0.19 & 0.50 & 0.17 & -69.9 & 35.4 & 0.39 & 0.38 & 30.3 \\
& & C4 & 15 & 0.08 & 1.25 & 0.19 & -57.0 & 15.5 & 0.67 & 1.00 & ... \\
& & C3 & 15 & 0.03 & 1.97 & 0.23 & -41.7 & 9.2 & 0.60 & 1.00 & ... \\
& & & 8 & 0.03 & 2.08 & 0.57 & -28.1 & 15.2 & 1.25 & 1.00 & ... \\
& & C1 & 8 & 0.02 & 7.39 & 0.55 & -31.7 & 4.5 & 2.39 & 1.00 & ... \\
& & & 2 & 0.06 & 7.18 & 1.95 & -32.3 & 16.5 & 4.50 & 0.37 & 18.6 \\
& 12 & C0 & 8 & 1.08 & ... & ... & ... & ... & 0.18 & 0.73 & -70.4 \\
& & C5 & 8 & 0.32 & 0.41 & 0.18 & -70.2 & 31.3 & 0.33 & 1.00 & ... \\
& & C4 & 8 & 0.21 & 1.34 & 0.17 & -53.8 & 10.8 & 0.79 & 0.49 & -41.8 \\
& & C3 & 8 & 0.03 & 2.09 & 0.17 & -48.4 & 6.9 & ... & ... & ... \\
& & C2 & 2 & 0.07 & 4.69 & 0.63 & -37.6 & 10.4 & 0.81 & 1.00 & ... \\
& & C1 & 2 & 0.15 & 8.49 & 0.62 & -42.0 & 5.9 & 1.35 & 1.00 & ... \\ \tableline
\end{tabular}}
\tablenotetext{a}{Epoch identification in Table~\ref{obslog} or Table~\ref{vlba}.}
\tablenotetext{b}{Flux density in Janskys.}
\tablenotetext{c}{$r$ and PA are the polar coordinates of the
center of the component relative to the presumed core C0.
Position angle is measured from north through east.}
\tablenotetext{d}{$\sigma_{r}$ and $\sigma_{{\rm PA}}$ are the
estimated errors in the component positions.}
\tablenotetext{e}{$a$ and $b$ are the FWHM of the major and minor axes of the Gaussian
component.}
\tablenotetext{f}{Position angle of the major axis measured from north through east.}
\end{table}

In Table~\ref{mfittab}, we have only included those components which we can unambiguously resolve.
The lower resolution 2 GHz images usually see the core plus the inner 8 GHz 
components as only a single component.  Since it is impossible to deconvolve these
merged components and know how much flux is coming from each, we do not
include these merged 2 GHz components in Table~\ref{mfittab}. 
In the model fits listed in Table~\ref{mfittab} we have fixed the location of the presumed
core to be at the origin and give positions for all other components relative
to the core.  For all three of these sources we have selected as the core the
brightest compact component at the end of the extended jet structure.   
In all of these sources this component also has the flattest spectral index. 

Determining accurate errors in the model-fit parameters is problematic.
Formal methods for calculating the errors, such as that described by
Biretta, Moore, \& Cohen (1986), rely on varying a given parameter
until the reduced $\chi^{2}$ increases by a certain factor.  This gives 
an estimate for the uncertainty in a parameter within that particular
observation, but it does not take into account the differences in
$(u,v)$ plane coverage among observations in a series, which may cause
differences in the model-fit parameters among observations which are
much larger than the associated formal errors.  
Since geodetic VLBI observations
can have drastically different antennas and $(u,v)$ plane coverages, we
might expect that this effect would dominate for a series of geodetic
VLBI images.  Indeed, we have used the method of Biretta, Moore, \& Cohen (1986)
to derive formal errors for the model-fit parameters of 1611+343 presented
in Paper I, and it is obvious from the large scatter in the points that
the formal errors are much too small.  

The only way to derive an accurate 
estimate for the errors is to model fit a series of images made from
observations with varying $(u,v)$ plane coverages over a small enough
time interval that the source structure does not change and note the
scatter in the model-fit parameters.  Many of the sources in
the USNO database have observations close enough together in time that this
can be done.  Since we expect errors in the component positions to be
proportional to the beam size, we express the errors as a fraction of the
beam and note that for nearly simultaneous images of 1611+343 (Paper I),
and 1606+106, the scatter in the model-fit component positions is about a quarter
of a beam FWHM.  We have accordingly used a quarter of a beam as the error
in the position for most of the components, except for those noted below. 
For the beam size in a given direction
we have used the maximum projection of the beam onto a line in that direction.

It is reasonable to suppose that model fitting will more accurately locate
bright, compact components.  From Condon (1997)
it can be derived that the positional error should be proportional to
$B^{-1/2}$, where $B$ is the surface brightness of the component.
We have accordingly used smaller error bars for the brighter, more compact
components.  We have used  1/6 of a beam for component C2 in 0202+149, and
1/16 of a beam for the very bright, very compact component C3 in CTA 26.
The bright 2 GHz component C1 in 0202+149 is elongated perpendicular to the radial
direction; we have used 1/10 of a beam for the radial error and 1/6 of a beam
for the angular error.  In all cases the observed scatter of the
positions about the fitted lines agrees with the chosen error bars.
For the fitted fluxes the calibration error must be added to the model
fitting error.  Since the calibration errors are relatively large for
these geodetic VLBI experiments, the model-fit fluxes are in
general not very accurate, except possibly for the very bright components.

\section{Motion of Components}
\label{motion}
In this section we discuss the motions of the individual components in both radius
and position angle.  Two of the sources, CTA 26 and 1606+106, have components moving
outward at apparent superluminal velocities, while the components in 0202+149
are clearly subluminal.  We have calculated values for the apparent velocities
by performing least-squares fits to the component distances from the cores as a 
function of time, using the model-fit positions and errors discussed in the
preceding section.  In all cases a linear fit was adequate, i.e. we have no
significant detection of acceleration or deceleration of any of the components.
When a component was detected at more than one frequency, we fit the positions measured at
each frequency separately to avoid effects from frequency-dependent separation.
We then took the weighted average of the velocities measured at the different
frequencies to obtain a single value for the velocity.
For all three of the sources discussed here,
the velocities measured for the individual components
are consistent with all components within the source having the same velocity.

Motion of the jet components in position angle is also discussed in this section.
All position angle values are measured from north through east.
Here we also present data on the motion of the components of 1156+295
in position angle, since only the outward motions of the 1156+295 components
were discussed in Paper II.  Curved jets on VLBI scales are a common feature
of blazars, and we observe bent jets in the sources we have studied.
In fact, we have highly significant detections of different position angles
for different components for all five of the sources we have studied
in detail.  Whether or not individual components move in position angle,
and whether or not they follow other components,
must be examined on a case-by-case basis.  This requires longer monitoring
of individual components, and preferably waiting until a component reaches
the same radial distance from the core previously occupied by an earlier component.
For 1611+343 (Paper I), we found that components were located at different
position angles, but that motion with constant position angle was adequate to
fit individual component motions.  We observed significantly different position angles for
components at the same radial distance, which ruled out motion of the components
along identical curved paths.

\subsection{0202+149}
The source 0202+149 is a flat-spectrum radio quasar with a redshift of 0.833 (Stickel et al. 1996).
Bondi et al. (1996), Padrielli et al. (1986), and Romney et al. (1984)
imaged this source with VLBI
at three epochs (1980 February, 1981 October, and 1987 November) at 1.7 GHz as
part of a campaign to study the structure of low-frequency variable sources.
They found that the structure of 0202+149 was well fit by a model with two components separated
by $\approx$3.4 milliarcseconds (mas) along a position angle of $\approx-71\arcdeg$.
The separation of the components did not change significantly over the
three epochs, although the low resolution of the observations does not allow the authors
to place a very strict upper limit on the proper motion.  They find that both
components vary in flux, the southeastern component brightens, and the northwestern
component fades over the three epochs.  A 22 GHz VLBI observation by Moellenbrock
et al. (1996) measured a brightness temperature for 0202+149 in excess of the inverse Compton
limit for synchrotron radiation (10$^{12}$ K), indicating the likelihood of relativistic
beaming in this source.

The separations from the core as a function of time that we measure from our images
are shown in Figures 7$a$ and $b$ for components C1 and C2 respectively.
By taking the weighted average of the velocities found from separate
fits to the 2, 8, and 15 GHz positions of
C1, we calculate an outward velocity of 0.11$\pm$0.88 $h^{-1}c$ for this component.
Similarly, we measure a velocity of $-$0.23$\pm$0.65 $h^{-1}c$ for component C2.
Both of these velocities are consistent with no motion, and the one-sigma
upper limits are 0.99 and 0.42 $h^{-1}c$ for C1 and C2 respectively.
Since the measured positions are consistent with stationary components, we have
plotted the best fits to constant separation from the core in Figures 7$a$ and $b$.
In Figure 7$a$ the fitted separations are at successively greater
distances from the core as the frequency increases.  This frequency-dependent 
separation was also seen in 1156+295 (Paper II) and has been noted by other authors
(e.g. Biretta, Moore, \& Cohen 1986).  It may be
due to gradients in magnetic field and electron density, which cause
the $\tau =1$ surface to move progressively inward at higher frequencies.
The frequency-dependent separation in Figure 7$b$ appears to go in the wrong
direction, although the error bars on the 15 GHz points are such that this
separation is not significant.

\begin{figure}
\plotfiddle{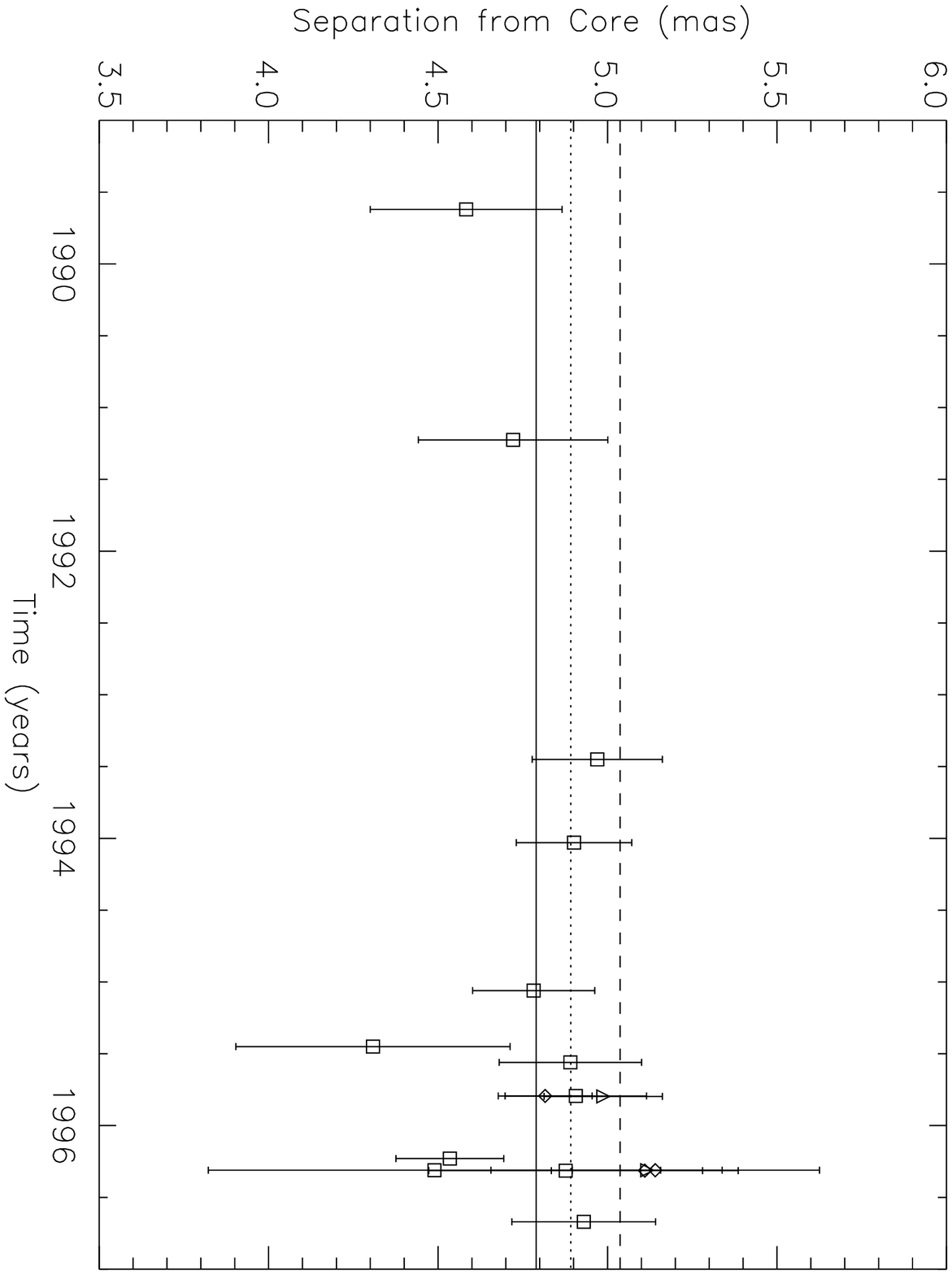}{7.4 in}{180}{82}{82}{482}{607}
Figure 7$a$: Separation of component C1 in 0202+149.
Squares represent the 2 GHz positions,
diamonds 8 GHz, and triangles 15 GHz.
The lines shown are the best fits
to constant separation from the core.  The solid line is the best fit to the 2 GHz
positions, the dotted line to the 8 GHz positions, and the
dashed line to the 15 GHz positions.
\end{figure}
\begin{figure}
\plotfiddle{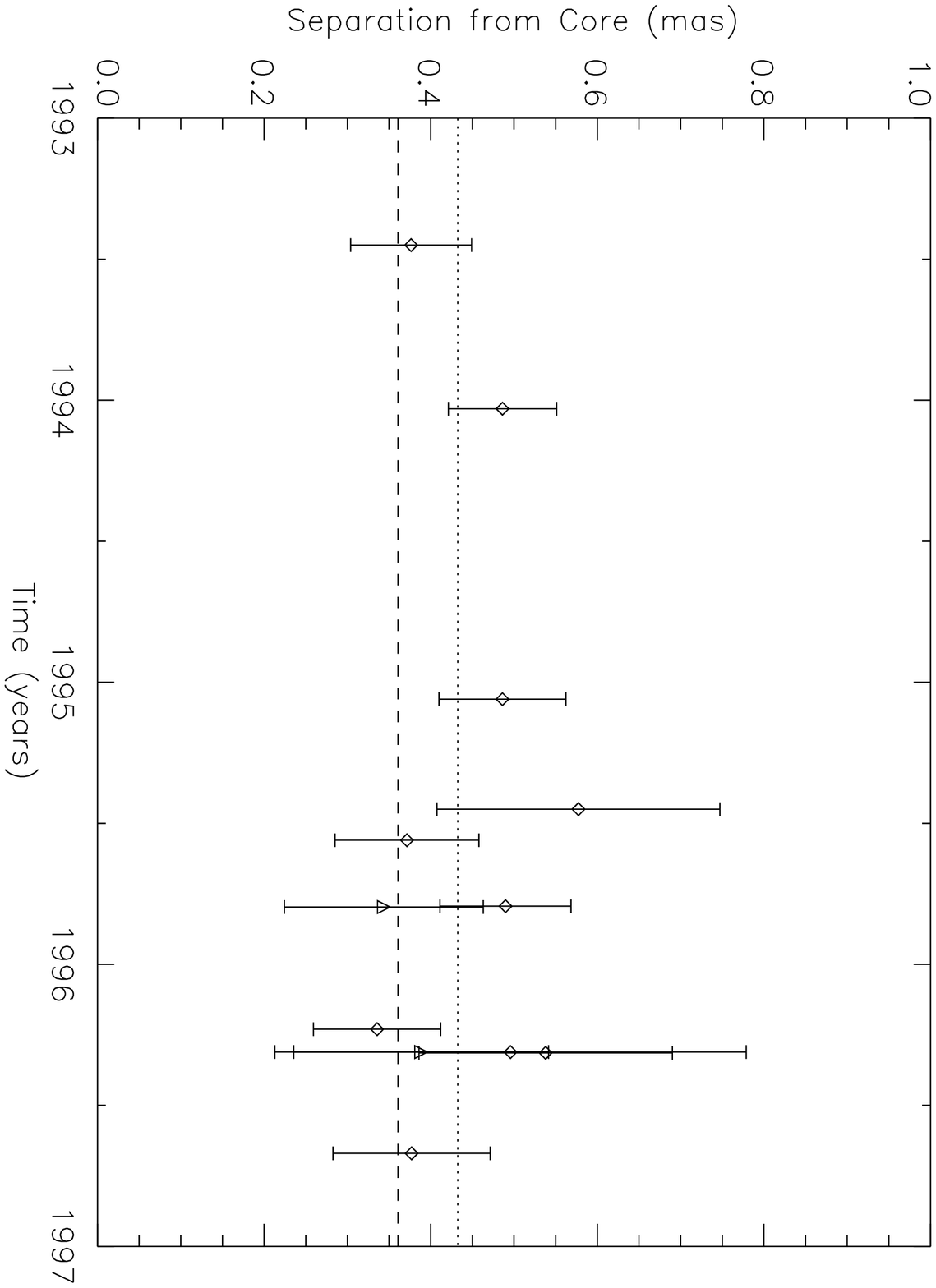}{7.4 in}{180}{82}{82}{482}{607}
Figure 7$b$: Separation of component C2 in 0202+149.
Diamonds represent the 8 GHz positions and triangles the 15 GHz positions.
The lines shown are the best fits
to constant separation from the core.  
The dotted line is the best fit to the 8 GHz positions, and the
dashed line to the 15 GHz positions.
\end{figure}

The VLBI observations of 0202+149 at 1.7 GHz presented by Bondi et al. (1996) show
a bright component about 3.4 mas from the core which can probably be identified
with our component C1.  Although their three measured positions are also consistent
with the component having no motion, their measured separation is
about 1.4 mas closer to the core than ours.  If a fit is done to the
Bondi et al. (1996) positions which cover the time range 1980 to 1987,
combined with our measured 2 GHz positions which cover the time range 1989 to 1996, 
then a slightly superluminal velocity of 1.8$\pm$0.5 $h^{-1}c$ is obtained.
However, this velocity comes entirely from the difference in separation
between the Bondi et al. (1996) points and our points, and each set of
measurements individually is consistent with no motion.  
We suspect that this velocity may arise from frequency-dependent separation combined
with systematic differences between the two sets of observations, 
and we use only the velocity obtained from fits to our actual measured
component positions.

Component C2 also appears to be stationary at 0.41 mas from the core, 
but this component is not present
in the images from 1989 and 1991.  Either this component really is moving 
outward, or it may represent a standing shock in the underlying flow which
appeared in this location sometime between 1991
and 1993.  If this component were moving at its 2$\sigma$ upper limit proper
motion of 0.04 mas yr$^{-1}$, then in 1991 it could have been only 0.33 mas
from the core and may have been unresolved in these earlier images.  The high
8 GHz core flux in the 1991 image may support this.  If this is so, then the
velocity of C2 would actually be around 1.0 $h^{-1}c$.

The components in this source are stationary in position angle as well as in radius.
However, they have different position angles from each other.  The measured
position angle of C1 is $-53.0\pm0.9\arcdeg$, and that of C2 is
$-75.0\pm4.3\arcdeg$.  The average position angle of C1 from the
observations presented by Bondi et al. (1996) is $-71\arcdeg$, with no
error bars given for the position angles.  Their error bars are likely
to be quite large because of the large size and ellipticity of their beams.
If the difference between the Bondi et al. (1996) C1 position angle and our C1
position angle is in fact significant, this would imply motion of C1 from
a position angle close to that of C2 to its present
position angle over the time between the Bondi et al. (1996) observations and the present.
This would imply motion of C1 at close to its one-sigma upper limit
velocity.

\subsection{CTA 26}
The quasar CTA 26 (0336-019) is a core-dominated flat-spectrum radio source at a
redshift of 0.852 (Hewitt \& Burbidge 1989).  A 5 GHz VLBI image from
Wehrle et al. (1992) shows a jet at a position angle of about 65$\arcdeg$,
roughly orthogonal to the VLA secondary.  Relativistic beaming is indicated by the
VLBI observations of Linfield et al. (1989), who measure a brightness temperature
well in excess of the 10$^{12}$ K inverse Compton limit.

Figures 8$a$ and $b$ show the outward motions that we measure for the components of CTA 26.
The motion of the outermost component C1 is shown in Figure 8$a$.  This 
component was only detected at 2 GHz.  The lower 2 GHz resolution, in
combination with very elliptical beams at some epochs, has resulted in
rather large errors in the position measurements for this component.  These
errors lead to a large error in the fitted velocity, which
is 12.6$\pm$9.6 $h^{-1}c$.  This component fades over time and is not
visible in the 2 GHz images after mid 1995.

\begin{figure}
\plotfiddle{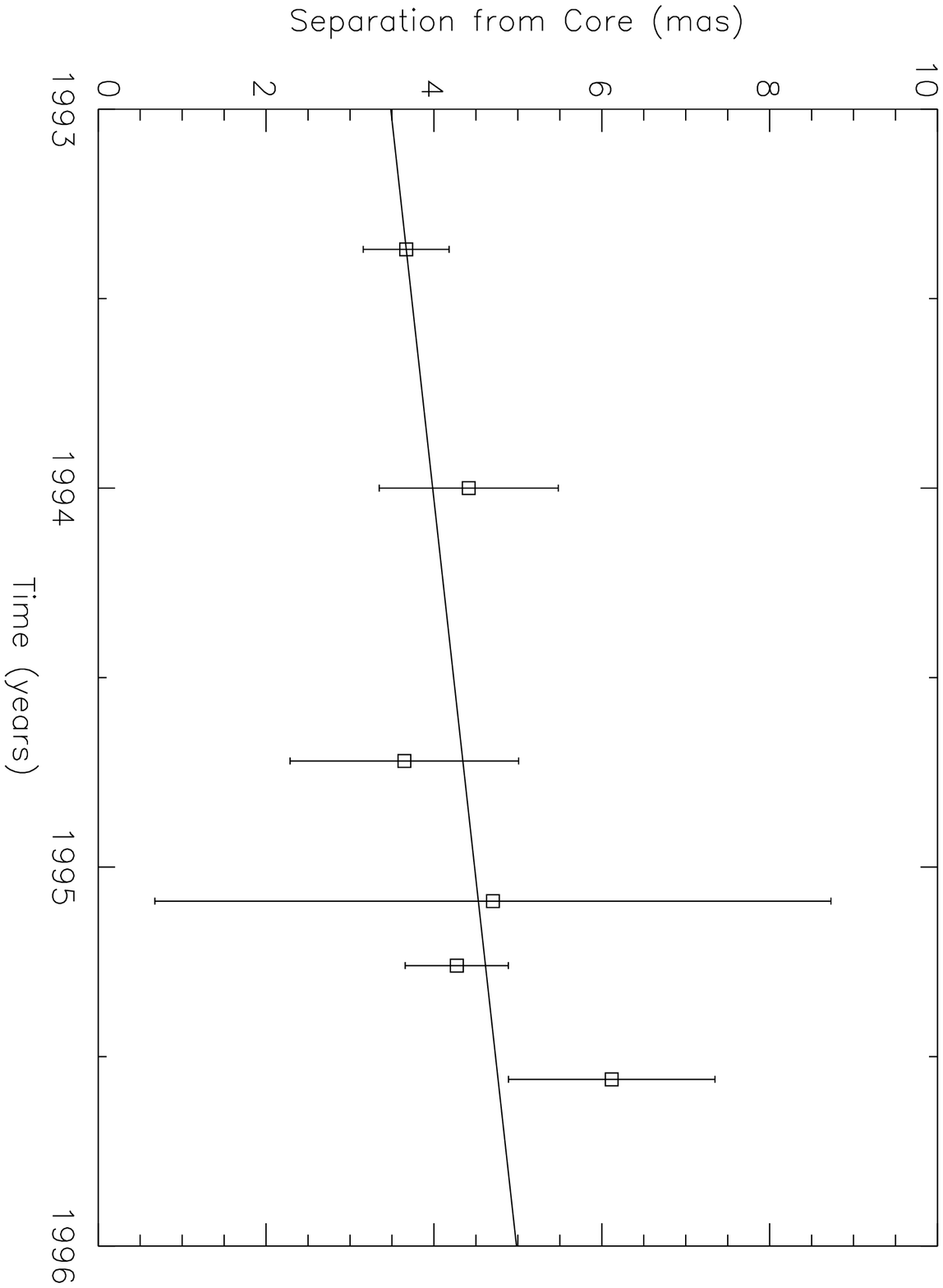}{7.4 in}{180}{82}{82}{482}{607}
Figure 8$a$: Separation of component C1 in CTA 26.
All measurements are at 2 GHz. 
The line is the best fit
to motion with constant velocity.
\end{figure}
\begin{figure}
\plotfiddle{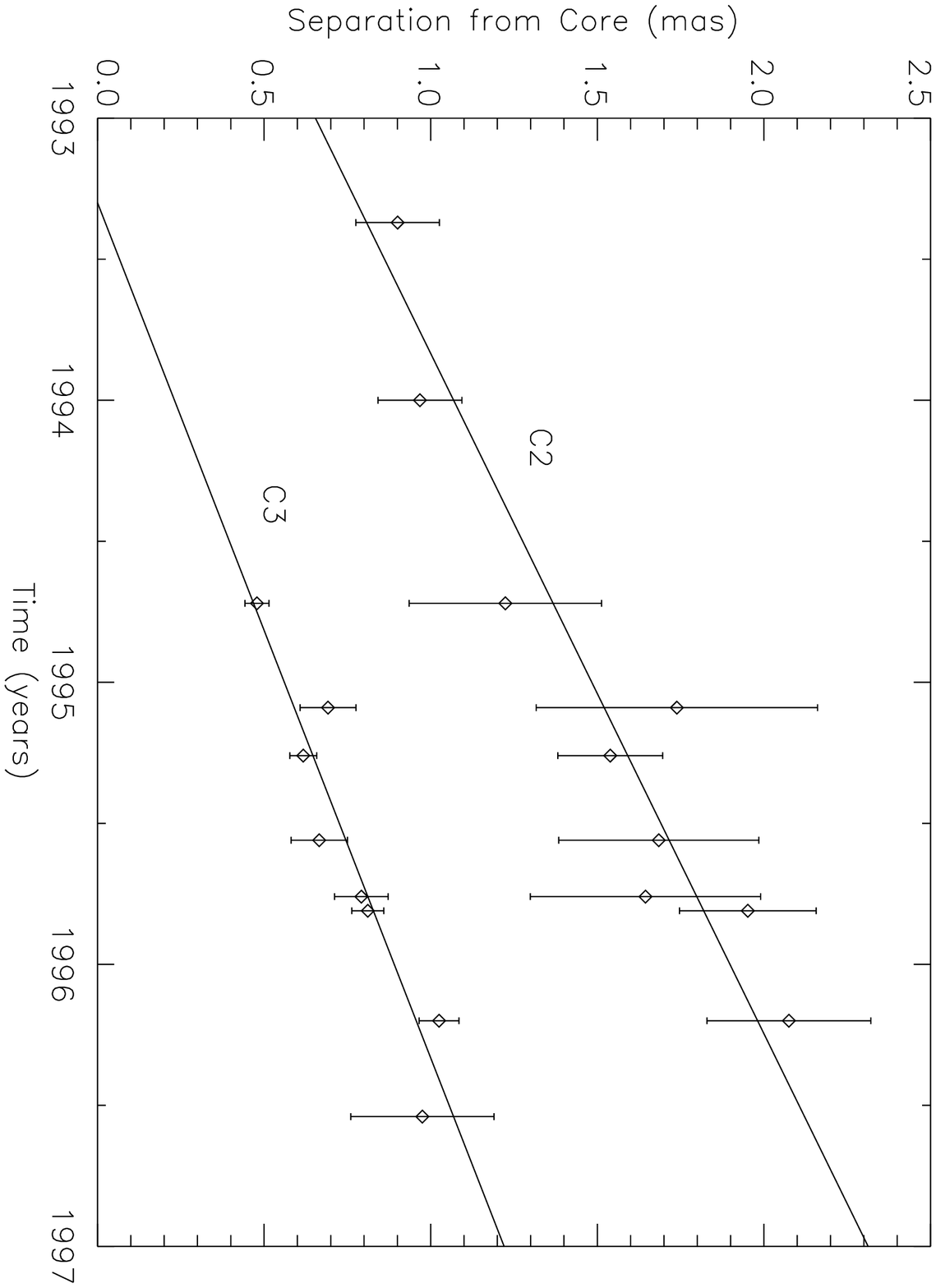}{7.4 in}{180}{82}{82}{482}{607}
Figure 8$b$: Separation of components C2 and C3 in CTA 26.
All measurements are at 8 GHz.
The lines are the best fits
to motion with constant velocity.
\end{figure}

The motions of the two components which we detected in the 8 GHz
images, C2 and C3, are shown in Figure 8$b$.  Component C2 was detected
in the earliest image and continued to be visible over most of the
covered time range, although it faded as time went on.  Component C3
was first detected in late 1994 and was visible in all subsequent images.
The measured outward velocities for these two components are 10.5$\pm$1.6 $h^{-1}c$
for C2, and 8.3$\pm$1.0 $h^{-1}c$ for C3.
These measured velocities and standard errors are consistent
with all three components moving at the weighted average velocity of 8.9$\pm$0.8 $h^{-1}c$.
This value is near the high end of the apparent velocity distribution for core-dominated
quasars.  This distribution appears to have a cutoff around speeds of 10 $h^{-1}c$
(VC94); however, velocities between 8 and 10 $h^{-1}c$ are not
uncommon. 

Figure 9 shows the measured separations and position angles
of all components at all frequencies
and all epochs for CTA 26.  We plot $r$ on a logarithmic scale to aid in
display of both inner and outer components.  Motion with constant position
angle corresponds to motion along a horizontal line on this plot.
The motion of C3 is well fit by motion at a constant position angle
of $66.6\pm2.0\arcdeg$; however,
we have a greater than 2$\sigma$ detection of change in position angle
for both C2 and C1.
C2 decreases its position angle from about $100\arcdeg$ to about $60\arcdeg$
as it moves from 1 to 2 mas, while C1 increases its position angle
with time.  If C2 and C1 define a single path, evidently the
components move toward lower position angles from radii around 1 mas to
radii around 3 or 4 mas, at which time they curve back to higher position angles.
Such oscillations in position angle are expected in the helical models which
have been proposed to explain some curved jets (e.g. Hardee 1987).
Although C2 and C1 may define a single path, it is evident that C3 has taken
a different path out from the core than C2.  C3 is at a lower position angle
than C2 when C2 was at the same radius.  If C3 continues moving along its
present straight path, it will merge with the path defined by C2 at a radius
of between 1 and 2 mas.  Components which take different paths out from the
core and later merge onto the same curved path have been seen in 3C345
(Zensus, Cohen, \& Unwin 1995).

\begin{figure}
\plotfiddle{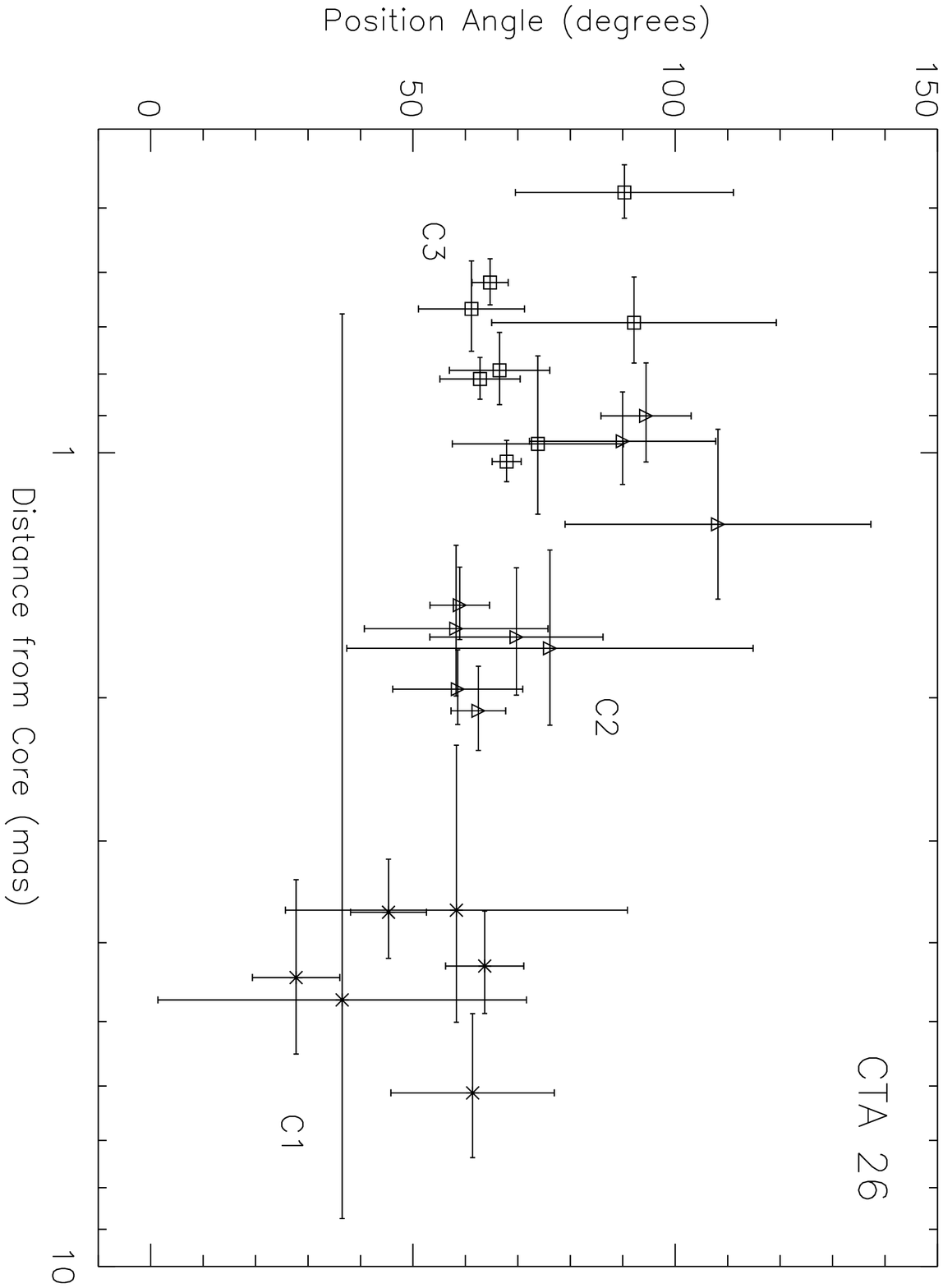}{7.4 in}{180}{82}{82}{482}{607}
Figure 9: Measured positions of all components at all frequencies and all epochs for CTA 26.
The $x$-axis shows the distance from the core on a logarithmic scale and
the $y$-axis shows the position angle, measured from north through east.
Asterisks show measured positions of C1, triangles C2, and squares C3.
\end{figure}

\subsection{1156+295}
The geodetic VLBI images and outward motions of the components of 1156+295 were 
discussed in Paper II.  Here we discuss the motion of the components of this
source in position angle.  Figure 10 is the same as Figure 9 except it shows the component
positions from 1156+295.  The 2 GHz positions have been shifted out
by 0.65 mas to correct for the observed frequency-dependent separation (Paper II).
The motions of C1, C2, and C4 are well fit by constant position angles,
which differ significantly from each other, of $36.0\pm2.6\arcdeg$, $15.2\pm1.9\arcdeg$, and
$9.7\pm6.0\arcdeg$ respectively.  We have a greater than 2$\sigma$ detection of a
change in position angle for C3, from $40\arcdeg$ at 1 mas to $-20\arcdeg$ at 3 mas.
Inspection of Figure 10 suggests a continuous path, consisting of oscillations in position angle,
that the components could follow.  The components could start at the core with
a position angle of $0\arcdeg$ and increase their position angle to about
$20\arcdeg$ at 1 mas, at which point they would turn and move toward lower
position angles of around $-20\arcdeg$ at 3 mas.  They could then curve back toward
higher position angles, reaching a position angle of $35\arcdeg$ at 10 mas.
Whether or not such a path exists should become evident with continued monitoring.
C4 has already reached the earliest C3 radius and is somewhat below that C3
position angle; however, the error bars on the earliest C3 point are large enough
that this may not be significant.  C3 has just now reached the radius of the
earliest C2 point.  If it is going to follow C2, it needs to start moving toward
higher position angles soon.

\begin{figure}
\plotfiddle{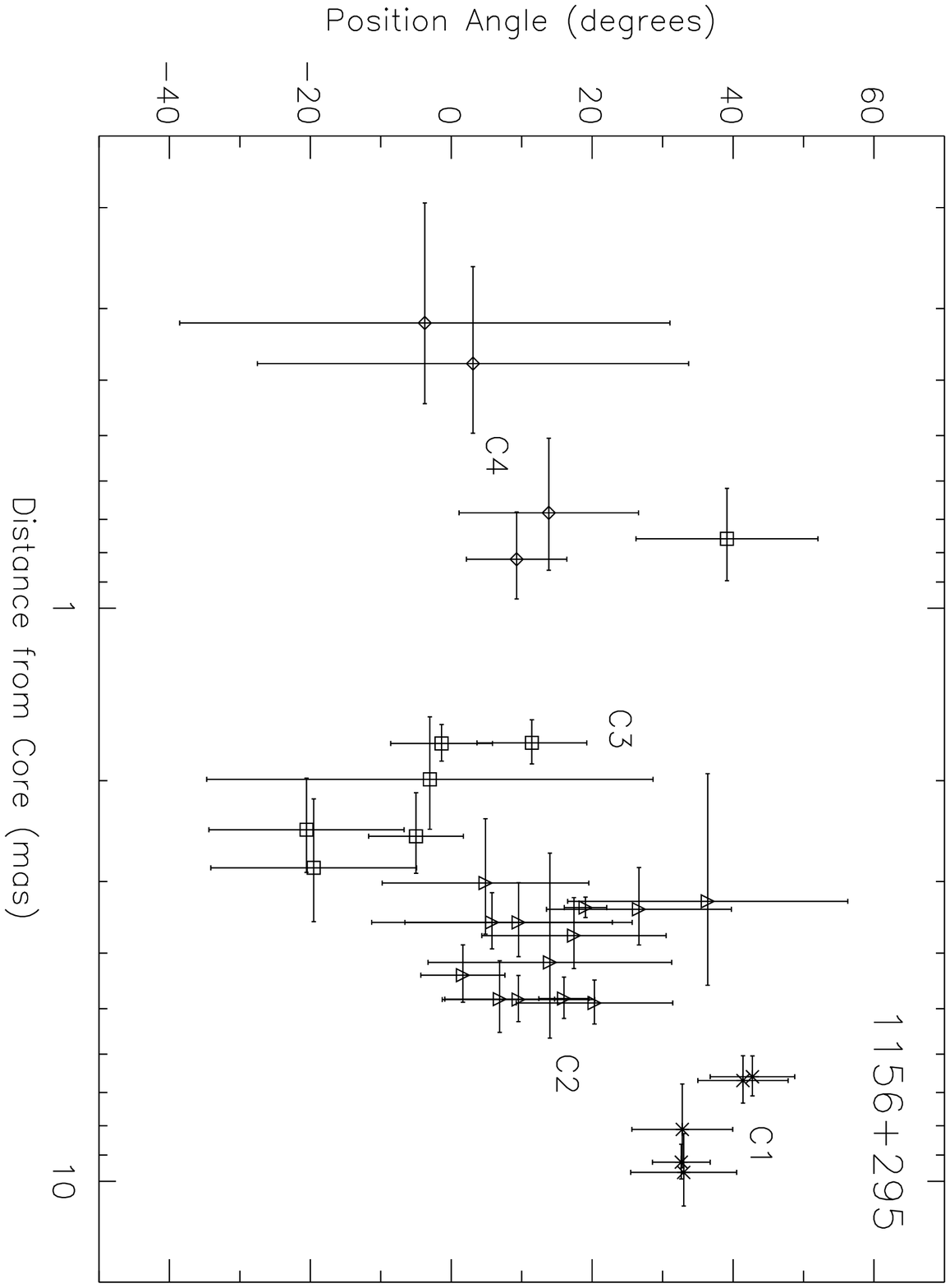}{7.4 in}{180}{82}{82}{482}{607}
Figure 10: Measured positions of all components at all frequencies and all epochs for 1156+295.
The $x$-axis shows the distance from the core on a logarithmic scale and
the $y$-axis shows the position angle, measured from north through east.
Asterisks show measured positions of C1, triangles C2,
squares C3, and diamonds C4.
\end{figure}

\subsection{1606+106}
The flat-spectrum radio quasar 1606+106 has a redshift of 1.23 (Stickel \& K\"{u}hr 1994).
This quasar was observed with VLBI at 5 GHz in a three-antenna experiment by Zensus,
Porcas, \& Pauliny-Toth (1984).
It was well fit by a circular Gaussian of width 0.6 mas.
At least three images of this source have been made from geodetic VLBI data by
Britzen et al. (1994), although the images are not presented in that reference.

The components in the jet of 1606+106 are considerably fainter than the components
in the other two sources studied here, so not all components are detected by us at
all epochs.  In general, our later observations of 1606+106 had superior sensitivity
and $(u,v)$ plane coverage, and were better able to detect the fainter components.
Figure 11$a$ shows the measured positions of the outer components, C1 and C2, from our images.
These components were both mainly detected in the 2 GHz images, although there are some
8 GHz detections of each.  The weighted average of the velocities measured at 2 and
8 GHz are 8.0$\pm$5.9 and 6.1$\pm$6.0 $h^{-1}c$ for C1 and C2 respectively, where
the large error bars are the result of the lower 2 GHz resolution.  

\begin{figure}
\plotfiddle{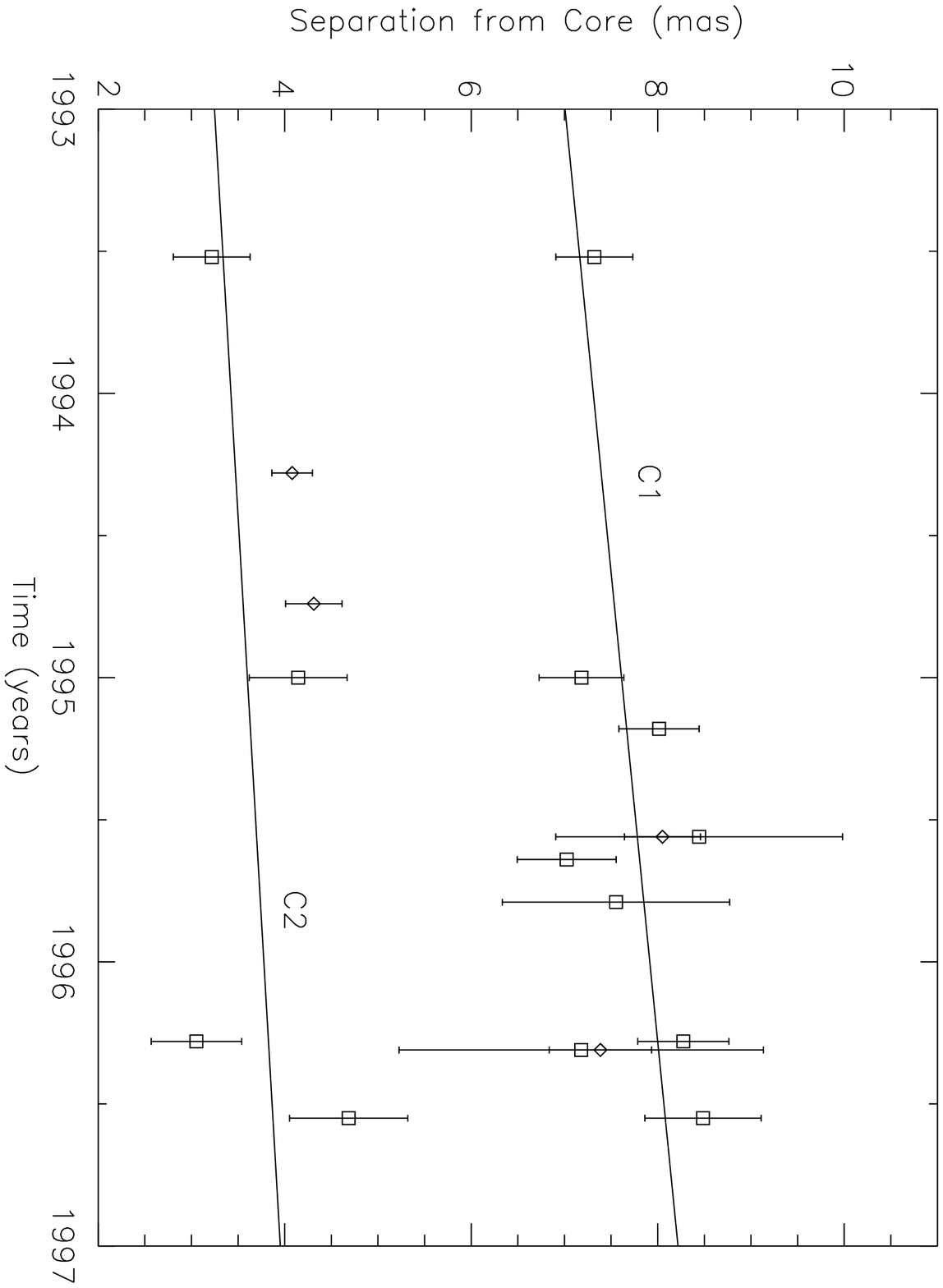}{7.4 in}{180}{82}{82}{482}{607}
Figure 11$a$: Separation of components C1 and C2 in 1606+106.
Squares represent the 2 GHz positions and diamonds the 8 GHz positions.
The lines are the best fits
to motion with constant velocity for the 2 GHz positions.
\end{figure}

Figure 11$b$ shows our measurements of the 
distances from the core of the inner components: C3, C4, and C5.
These components are all too close to the core to be detected at 2 GHz.  All of
our observations of these components are at 8 GHz and are combined with some VLBA detections
at 15 GHz.  The measured velocities for components C3, C4, and C5 are 3.4$\pm$1.7,
2.4$\pm$1.1, and 0.0$\pm$0.9 $h^{-1}c$ respectively.  C3 and C4 move outward
superluminally, while C5 is stationary with respect to the core.
VC94 point out that the large number of
stationary components found in core-dominated quasars precludes their being part
of a continuous distribution of Lorentz factors, and they probably represent
a different phenomenon from the more rapidly moving components.  If we therefore
exclude the velocity measurement of C5, the velocities of the other four components
are consistent with all of these components moving at the same average velocity
of 2.9$\pm$0.9 $h^{-1}c$. 

\begin{figure}
\plotfiddle{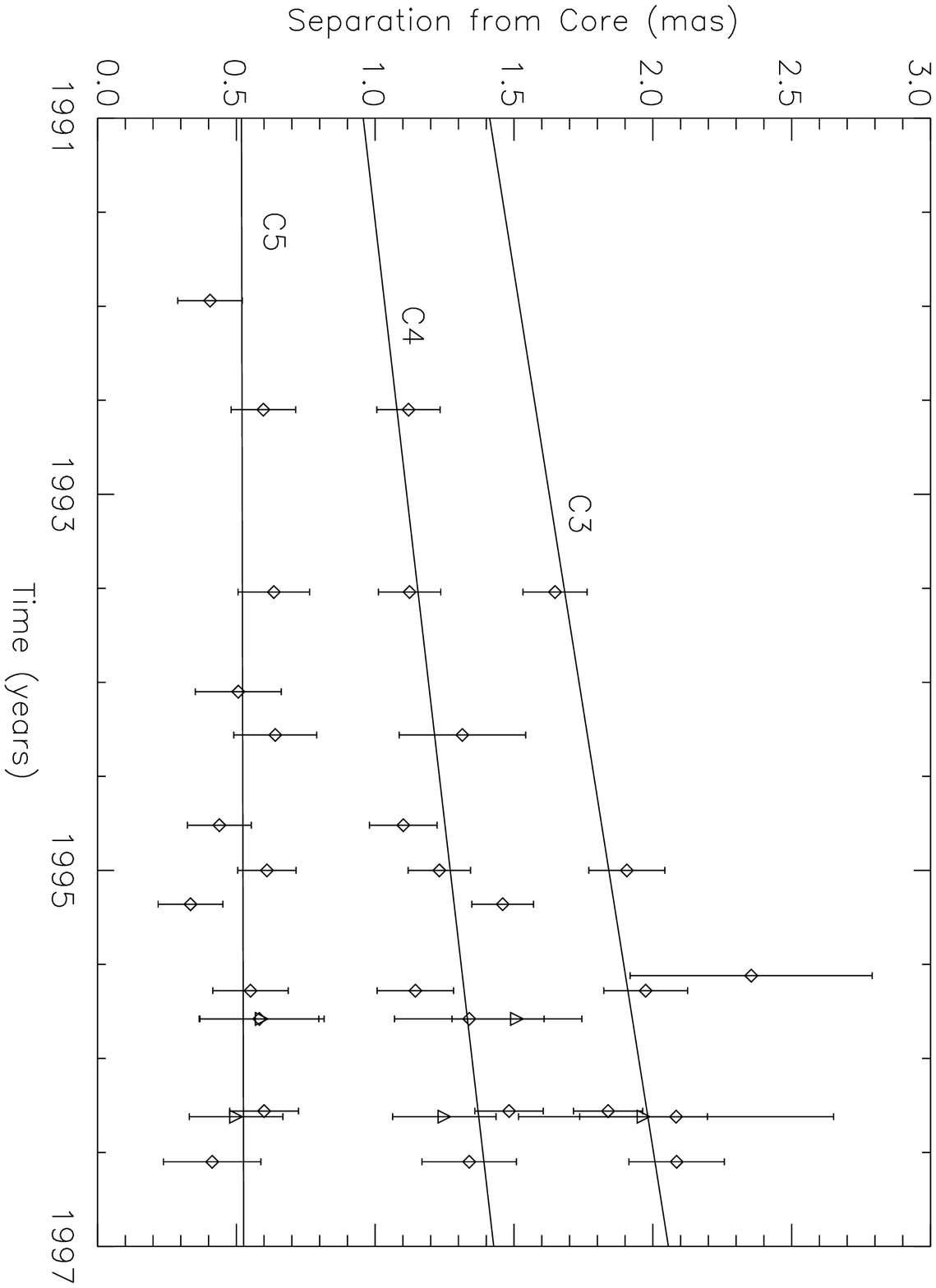}{7.4 in}{180}{82}{82}{482}{607}
Figure 11$b$: Separation of components C3, C4, and C5 in 1606+106.
Diamonds represent the 8 GHz positions and triangles the 15 GHz positions.
The lines are the best fits
to motion with constant velocity for the 8 GHz positions.
\end{figure}

We have more measured component positions for
this quasar than for any other; a total of 51 measured component positions
are plotted in Figure 12.  To aid comparison with the 8 GHz positions,
the 2 GHz and 15 GHz radii have been adjusted
by the observed magnitude of frequency-dependent separation in this source.
Each component's motion is well fit
by a constant position angle.  The motions of C1, C2, and C3 are all well fit
by motion along the same position angle, $-38.2\pm1.0\arcdeg$,
and the position angles of C4 and C5,
$-55.3\pm2.3\arcdeg$ and $-77.5\pm4.8\arcdeg$ respectively,
differ significantly from this and from each other.
In this source each component
could continue moving along the position angles given here, or C4 and
C5 could move toward higher position angles to join the path defined by C1,
C2, and C3.  If these component positions do define a continuous path,
then this path has only a gradual change in position angle which then
becomes constant around 2 mas, in contrast to the hypothetical paths in
CTA 26 and 1156+295, which had oscillations in position angle.

\begin{figure}
\plotfiddle{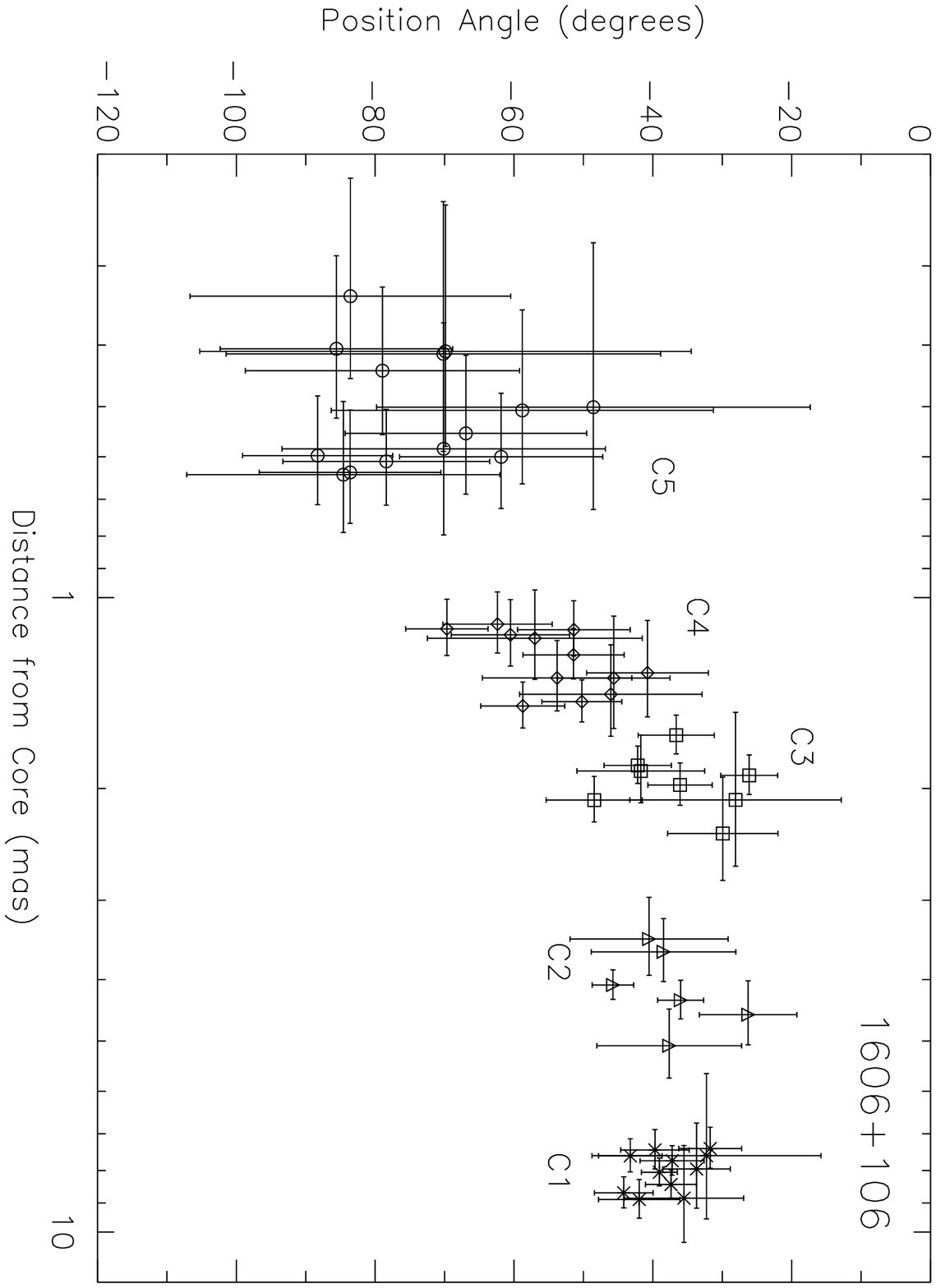}{7.4 in}{180}{82}{82}{482}{607}
Figure 12: Measured positions of all components at all frequencies and all epochs for 1606+106.
The $x$-axis shows the distance from the core on a logarithmic scale and
the $y$-axis shows the position angle, measured from north through east.
Asterisks show measured positions of C1, triangles C2,
squares C3, diamonds C4, and circles C5.
\end{figure}
 
\section {Discussion}
\label{discussion}
\subsection{0202+149 as a Compact F Double}
\label{compactf}
Analysis of the structure of VLBI images, along with large-scale source
properties, has enabled VLBI sources to be classified into a number of morphological
groups (e.g. Pearson \& Readhead 1988, hereafter PR88).
Most of the EGRET blazars imaged with VLBI
show the core-dominated, flat-spectrum, core-jet morphology characteristic of 
strong beaming and can be classified into the {\em very compact}, {\em compact},
or {\em asymmetric I} categories in the PR88 scheme.
All of the EGRET sources we have studied in detail have conformed to
this expected morphology, with
the exception of 0202+149.  This source distinguishes itself in our
VLBI images by having a simple double morphology at 2 GHz and by being
the only source, out of the five for which we have measured apparent velocities,
which exhibits no apparent motion of the components.  This source is
unusual for an EGRET source in its large-scale properties as well.
It is unresolved in VLA images (Murphy, Browne, \& Perley 1993), 
has a low radio polarization of
0.2\% (Perley 1982), and has a relatively small radio variability 
(V=0.3 from the UMRAO online database\footnote{http://www.astro.lsa.umich.edu/obs/radiotel/umrao.html}, 
where V=$\Delta S/<S>$).  These properties
all match the properties of the compact F double class,
and we can identify 0202+149 as a new compact F double source.  The 
compact double sources were first identified by Phillips and Mutel (1982),
and this class of sources was divided according to their spectral index
into compact S ($\alpha <-0.5$) and compact F ($\alpha >-0.5$) doubles by PR88.

Detailed studies of the compact double sources by Conway et al. (1994)
have shown that the compact S and compact F doubles
are likely to be very different physical phenomena and that the compact F doubles
are probably physically more similar to the core-jet sources than to their
steep-spectrum counterparts.  They suggest that the components of compact S doubles
are the termination points of oppositely directed jets, while compact F doubles
are an unusual form of core-jet source with a bright, stationary jet component.
The EGRET detection of 0202+149 strengthens the conclusion of Conway et al. (1994) that the compact
F double sources are just an unusual form of core-jet source and are physically
very different from the ``terminating jet'' interpretation of 
compact S doubles, where we would not expect to see $\gamma$-ray
emission. 

Observations of $\gamma$-rays from 0202+149 imply that the source is strongly
beamed.  This strong beaming, together with the very low apparent velocities, means
that the jet must be aligned extremely close to the line-of-sight if the component
velocities reflect the bulk motion of material in the jet.
Indications of strong beaming have been seen in the other compact F doubles in the
form of high jet/counterjet brightness ratios.  Conway et al. (1994) show
that the stationary components then imply such a small angle of the jet to the line-of-sight
that the probability of observing these sources is very low.
They suggest two models to explain the stationary components of
compact F doubles, although the large-scale properties of these sources, which
also make them quite different from normal core-jet sources, remain to be explained.  
The components could represent standing shocks in the jet, implying
that the component velocities are much less than the bulk velocity of jet material.
Alternatively, the components could be due to relativistic flow close to the line-of-sight
along a curving jet, and since a curving jet samples a range of angles to the line-of-sight,
the arguments suggesting a low probability of detection would not apply.
Since both of these models allow for bulk relativistic motion close to the line-of-sight,
the detection of $\gamma$-rays from a compact F double seems compatible with either model.

\subsection{Correlations between Flares and Component Ejections}
\label{flares}
It is interesting to see if VLBI component ejections correlate with outbursts
in the light curves of these sources.  Correlations between
the emergence of new VLBI components and radio flares have been noted by
many authors, e.g, Mutel et al. (1994, 1990), Zensus et al. (1990).  
Such a correlation is not surprising;
since newly emerged VLBI components typically have a high radio flux,
it is not unexpected that they would contribute significantly to the total light curve.
A more meaningful correlation would be if component ejections correlated with 
flares in the optical or $\gamma$-ray regimes.  Then we would actually be observing
the effects of enhanced activity in the central engine as they propagated down
the jet.  In the common shock interpretation of VLBI components, $\gamma$-ray and
optical emission could originate in these compact energetic shocks as they form
and move out from the core.  After some time, typically zero to several hundred days (Tornikoski et al. 1994),
the component becomes optically thin to lower frequency radio emission,
causing the observed brightening
in the radio.  Also at about this time, depending on source distance and VLBI resolution, the component
becomes resolved from the ``core'' in VLBI observations, where the VLBI ``core'' is
simply the point at which the jet becomes optically thick.
Such a detailed scenario in which the $\gamma$-ray emission is due to
a propagating discontinuity (corresponding to a jet
component observed with VLBI) in a Poynting flux jet, 
and where the delay of the radio emission is due to self-absorption or
free-free absorption by external plasma, has been proposed by Romanova \& Lovelace (1996).
Reich et al. (1993) have observed enhanced radio emission 
with a delay of several months relative to the EGRET detection
in some sources.  Conversely, some models predict
that there should be no correlation between VLBI component ejections and $\gamma$-ray flares.
In the model of Punsly (1996), the VLBI jet and the $\gamma$-ray jet are two separate jets,
so we would not expect any correlation between them.   

Correlations between optical flares and the formation of VLBI components have been noted ---
for example, component C9 of 3C273 (Krichbaum et al. 1990) and components K1 and K3 of OJ287
(Vicente et al. 1996) --- although these references also mention that there are optical flares
with no subsequent VLBI components and VLBI components which have no associated optical flares.
Gabuzda \& Sitko (1994) see variations in optical
polarization accompanying component ejections in OJ287 and 3C279.
The lack of detection of a VLBI component following a flare could be due to extremely short-lived
components.  Abraham et al. (1994) have detected components with lifetimes as short as
one year in 3C273, and they suggest the ejection rate of these components may be much higher
than previously thought.  
High frequency VLBI observations by Krichbaum et al. (1995) have revealed newly emerged 
components which appear correlated with preceding EGRET flares in 0528+134, 0836+710, and 3C454.3.
Similar results have been obtained for 3C279 (Wehrle et al. 1996).
These results indicate that $\gamma$-ray flares may be related to the production of observed
jet components.

We have calculated the epoch of zero separation for each moving component in all of our sources.
These times are listed in Table~\ref{ejtime}.  
We list only those components that have emerged recently
enough that correlated flux observations might exist, i.e., the outer components
of all sources and the slow-moving and stationary components of 0202+149 and 1606+106
are not listed.  The 1$\sigma$ errors in the velocities  
were used to calculate the errors in the separation times; 
note that this does not produce symmetric errors.
The recent ejections of C2 and C3 in CTA 26, and C3 and C4 in 1156+295, appear
to correlate with outbursts in the historic radio and optical light curves.  These outbursts
are listed in Table~\ref{ejtime}.  
For CTA 26 the ejection times of C2 and C3 correlate with the two
largest radio flares since 1990 (M. Aller 1996, private communication).  Smith et al. (1993) also
show CTA 26 entering a more active optical state around 1990.  For 1156+295, the ejection of C3
seems to correlate with a very large optical flare in 1985, and the later ejection of C4
corresponds to a peak in the high-frequency radio light curve (Tornikoski et al. 1994).

\begin{table}[!h]
\caption{Component Ejection Times}
\label{ejtime}
\begin{tabular}{c c c c c c} \tableline \tableline
& & & & Possible Correlated & \\
Source & Comp. & Ejection Time & 1$\sigma$ Time Range & Outbursts \\ \tableline
CTA 26 & C2 & 1991 Jun & 1990 Nov - 1991 Nov & 1990; 4.8, 8, and 14.5 GHz \\
& C3 & 1993 Apr & 1993 Jan - 1993 Jul & 1993; 4.8, 8, and 14.5 GHz \\
1156+295 & C3 & 1985 Sep & 1984 Jul - 1986 Jul & early 1985; optical \\
& & & & mid 1985; 90 GHz \\
& C4 & 1991 Mar & 1989 Jun - 1992 Jan & 1990; 37 and 90 GHz \\ \tableline
\end{tabular}
\end{table}

We are particularly interested in possible correlations with EGRET flares.  We found in Paper I
that the separation time of C5 in 1611+343 was during a high state of $\gamma$-ray activity.
For the sources studied in detail in this paper, a $\gamma$-ray flare can be identified 
in the EGRET light curves of CTA 26, 1156+295, and 1606+106.
Figure 13 shows the EGRET light curves for these three sources along with that
for 1611+343, which is shown for completeness. 
These light curves have been constructed from the data in Mukherjee et al. (1997).
All of these sources have a $\gamma$-ray variability index greater than 1 (Mukherjee et al. 1997),
indicating strong variability.  These plots also show the time ranges for the
ejections of VLBI components which have emerged during the lifetime of EGRET.
The only component whose ejection time correlates
with an observed high level of $\gamma$-ray activity is component C5 of 1611+343.
EGRET upper limits recorded during the ejection times of the other components
do not necessarily mean there was no correlated flare, because such a flare could have occurred
when EGRET was not observing the source.  Actual evidence of no correlation must instead
come from EGRET flares with no accompanying VLBI components.  As yet, no components have
been detected emerging as a result of the flares of CTA 26, 1156+295, and 1606+106.
Lower limits can be set on the time
at which such a component could have emerged by assuming it travels at the average speed of
components in that source and would be resolved by the time it reached 0.4 mas from the core
(the distance of C2 in 0202+149 and C5 in 1606+106).  These lower limits are shown as right-facing
arrows on the light curves of CTA 26, 1156+295, and 1606+106.  For CTA 26 and 1606+106 these
times are only slightly later than the observed $\gamma$-ray flare, 
so a somewhat slower than average component may yet be seen (in fact, such a component has been reported
recently in high-frequency VLBA observations of CTA 26 by Hallum et al. (1997)), but for 1156+295 it
appears there is definitely no component correlated with the 1993 EGRET flare.  If there
was a correlated component in this source, then it is likely that it was so short lived that
it was not detected.  These data give no indication
that EGRET flares are correlated with VLBI component ejections,
contrary to the correlations observed by other authors discussed earlier in this section.
However, since the observations by other authors were at higher frequency and resolution --- and
thus better able to detect short-lived components --- such a correlation may still be possible.

\begin{figure}
\plotfiddle{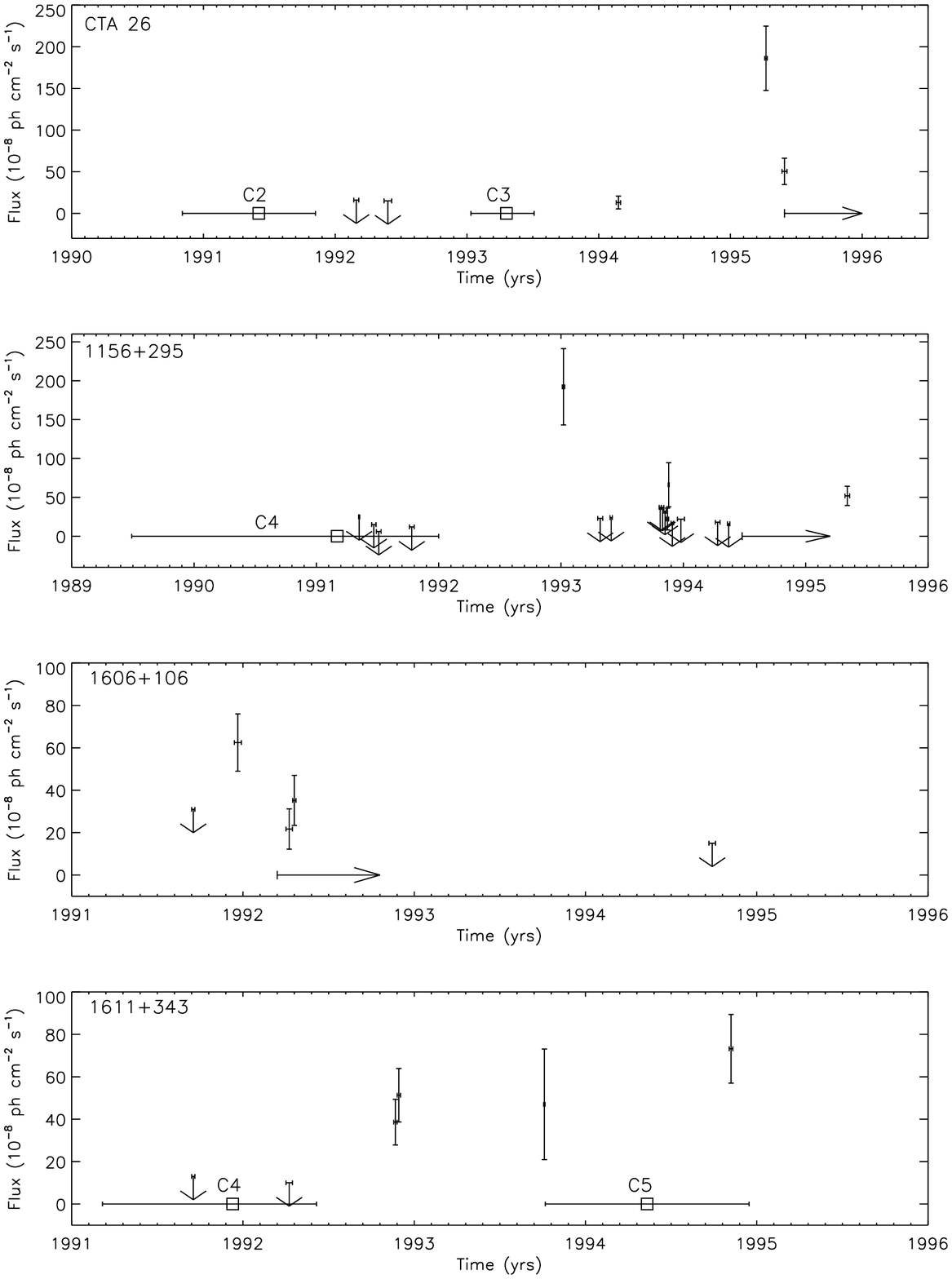}{7.625 in}{0}{80}{80}{-82}{-12}
Figure 13: EGRET light curves with component ejection times.
\end{figure}

\subsection{Calculation of Doppler Beaming Factors} 
Since the value of the apparent superluminal velocity is a function of both
the Lorentz factor $\Gamma$ and the angle to the line-of-sight 
$\theta$, knowledge of the apparent
speeds alone is sufficient only to place a lower limit on $\Gamma$ or an upper
limit on $\theta$.  Another parameter of the flow which depends on both 
$\Gamma$ and $\theta$ is the ratio of observed to emitted frequency, or
Doppler factor, $\delta =[\Gamma(1-\beta\cos\theta)]^{-1}$.  If both
$\delta$ and the apparent velocity are known, the Lorentz factor and
angle to the line-of-sight of the flow can in principle be determined.  
This assumes that the Lorentz factor of the pattern speed seen in the
superluminal motions, $\Gamma_{p}$, is the same as the Lorentz factor of
the bulk fluid flow causing the Doppler boosting, $\Gamma_{b}$, which for
some sources may be an incorrect assumption.

There have been several methods used to determine values of $\delta$ for AGNs.
One method, (see e.g. Marscher 1987; Ghisellini et al. 1993, hereafter GPCM),
assumes that the X-ray flux from the source is produced
by the synchrotron self-Compton (SSC) process.  The expected SSC emission from a VLBI
component depends on $\delta$ and on parameters which can be estimated
from the VLBI images.  A lower limit to the Doppler factor, $\delta_{SSC}$, is
obtained because for values of $\delta_{SSC}$ which are too low, the calculated X-ray flux from
a component is greater than the actual X-ray flux from the source.
Another method of calculating $\delta$ (e.g., Dondi \& Ghisellini 1995, hereafter DG95)
can be applied to $\gamma$-ray sources
by demanding that the optical depth due to pair-production be low enough for
$\gamma$-rays to escape.  This optical depth depends on $\delta$, so a lower limit,
$\delta_{\gamma}$, can be obtained by setting the optical depth equal to 1.
Readhead (1994) notes that intrinsic brightness temperatures of sources do not
vary much from the ``equipartition brightness temperature'' --- the brightness
temperature at which there is equipartition of energy between the radiating particles
and the magnetic field.  The observed brightness temperature can be compared
with the equipartition brightness temperature to determine the equipartition Doppler
factor, $\delta_{eq}$ (G\"{u}ijosa \& Daly 1996, hereafter GD96).  
Each of these methods has drawbacks;
the first two give only lower limits to $\delta$ and the third assumes equipartition,
which may not be valid in all sources.  A lower limit on the Doppler factor 
implies that when these 
Doppler factors are compared with apparent superluminal velocities, only lower limits
to $\Gamma$ and upper limits to $\theta$ can be obtained, although these limits are
often better than those obtained using the apparent velocity alone.

We have calculated $\delta_{SSC}$ for the core components of
the sources in our sample.  Unfortunately, these Doppler factors depend on powers
as high as the square of sometimes poorly known quantities such as the angular
size of the component and the frequency and flux at the peak of the synchrotron
spectrum.  Since we only have observations at two frequencies, we are not able
to precisely determine the peak of the synchrotron spectrum, and in general we 
assume a turnover at 8 GHz if the spectrum is rising over the observed range, a 
turnover at 2 GHz if it is falling, or a turnover at 5 GHz if it is flat.  We have
high resolution for each of our observing frequencies, which should give relatively
accurate measurements of the core angular size.  We also average the Doppler 
factors calculated at the individual epochs
to obtain the estimate of $\delta$ for each source. 
This averaging should alleviate some
of the error introduced by inaccurate measurements of angular size or flux density.
Although GPCM, DG95, or GD96 have calculated Doppler factors for all of our sources except
0202+149, they
all used single-epoch single-frequency VLBI data, and hence were forced to assume that the
frequency of observation was the turnover frequency.  We hope our multi-epoch 
dual-frequency data may provide better estimates of these Doppler factors.

We use equation (1) of GPCM to calculate $\delta_{SSC}$.  
This equation is \[\delta_{SSC}=f(\alpha)S_{m}\left[\frac{\ln(\nu_{b}/\nu_{m})\nu_{x}^{\alpha}}
{S_{x}\theta_{d}^{6-4\alpha}\nu_{m}^{5-3\alpha}}\right]^{1/(4-2\alpha)}(1+z),\]
where $f(\alpha)\approx-0.08\alpha+0.14$, $S_{m}$ is the flux density (in Jy) at the turnover
frequency $\nu_{m}$ (in GHz), $S_{x}$ is the X-ray flux density (in Jy) at frequency
$\nu_{x}$ (in keV), $\nu_{b}$ is the synchrotron high-frequency cutoff (assumed to be
$10^{14}$Hz, (GPCM)), $\theta_{d}$ is the angular diameter of the radio core at the turnover frequency
(in mas), and the sign of $\alpha$ has been changed to reflect the convention used in this paper.
This equation assumes the radiating component is a homogeneous sphere.  While
more sophisticated geometries such as conical jets have been used by some
authors, e.g. Unwin et al. (1994), they also require more sophisticated multi-frequency
VLBI data in order to constrain them.  In keeping with GPCM, we
use an optically thin spectral index of $\alpha=-0.75$ and use X-ray fluxes
from DG95.  GD96 report two correction
factors that must be applied to this equation: a factor of 1.8 in the angular size
and a factor of 2 in the flux at the spectral peak.  When we calculated Doppler factors
for 1611+343 in Paper I, we were aware of only the first of these correction factors,
so we give fully corrected values for 1611+343 here.  Values of the Doppler factors
and the corresponding limits to $\Gamma$ and $\theta$ are given in Table~\ref{doppler}.  Note
that since $\Gamma(\delta)$ has a minimum of 
$\Gamma^{*} =(\beta_{app}^{2} +1)^{1/2}$, where $\beta_{app}$ is the apparent velocity,
at $\delta =\Gamma^{*}$, then if $\delta <\Gamma^{*}$ the lower limit obtained on
$\Gamma$ is the same as that obtained from the apparent velocity alone. 

\begin{table}[!b]
\caption{Doppler Beaming Factors}
\label{doppler}
\begin{tabular}{c c c c c} \tableline \tableline
& & & & $\theta$\tablenotemark{c} \\
Source & $\delta_{SSC}$ 
& $\beta_{app}$\tablenotemark{a} & $\Gamma$\tablenotemark{b} & (deg) \\ \tableline
0202+149 & 18.0 & 0.4 & 9.0 & 0.15 \\
CTA 26 & 12.8 & 8.9 & 9.5 & 4.2 \\
1156+295 & 4.4 & 5.2 & 5.3 & 12.9 \\
1606+106 & 15.3 & 2.9 & 7.9 & 1.4 \\ 
1611+343 & 16.6 & 11.4 & 12.2 & 3.2 \\ \tableline
\end{tabular}
\tablenotetext{a}{Apparent velocity.  Values are the weighted average of the
apparent velocities of all components except for 1606+106, 
where the stationary component C5 is excluded,
and 1611+343, where the velocity of the inner component is
used since the components have different velocities.}
\tablenotetext{b}{Lower limit on the Lorentz factor calculated 
from $\delta_{SSC}$ and $\beta_{app}$.}
\tablenotetext{c}{Upper limit on the angle of the jet to the
line-of-sight calculated from $\delta_{SSC}$ and $\beta_{app}$.}
\end{table}

The values of $\delta_{SSC}$ found for CTA 26 and 1156+295 agree well with the
values found for these sources by GPCM and GD96.  The values of $\delta_{SSC}$
found for 1606+106 and 1611+343 are considerably higher than the corresponding
values found by DG95, possibly due to higher resolution of our VLBI observations.
The values of $\delta_{SSC}$ calculated for these five sources
are considerably higher than the values of $\delta_{\gamma}$ calculated for these
same sources by DG95.  In fact, the average value of $\delta_{SSC}$ for the
entire DG95 sample is higher than the average value of $\delta_{\gamma}$ for the
same sample, which implies that the $\gamma$-ray optical depth in these sources
is considerably less than 1, since a value of unity is assumed in the
calculation of $\delta_{\gamma}$.

The angles to the line-of-sight listed in Table~\ref{doppler} are all relatively small, as is
expected for strongly beamed sources.  In particular, the high Doppler factor and
low apparent velocity of 0202+149 imply a very small angle to the line-of-sight of
0.15$\arcdeg$.  The chance probability of a source having such a small angle to the
line-of-sight is only 3.4$\times10^{-6}$. 
Such an unreasonably small angle must
imply that the pattern speed $\Gamma_{p}$ is much less than the bulk fluid speed
$\Gamma_{b}$ (this equality being assumed in the calculation of $\Gamma$ and $\theta$),
or that the stationary components represent points along a curving jet where the
jet is directly aligned with the line-of-sight.  One of these situations is what we
would expect, given the nature of 0202+149 as a compact F double (see $\S$~\ref{compactf}).
A similar situation may exist for 1606+106, since it also has a low $\beta_{app}$ but high $\delta$.
The other sources also have angles close to
the line-of-sight, and in fact all except 1156+295 
are on the small angle side of
the maximum in the $\beta_{app}(\theta)$ relation, where the apparent velocity
{\em decreases} with decreasing angle to the line-of-sight.
If we consider a Hubble constant of 50 rather than 100,
thus doubling the values of $\beta_{app}$, the angles to the line-of-sight
become larger for all sources except 1156+295, 
and the angular solutions for CTA 26 and 1611+343 move to the large angle side of
the maximum in the $\beta_{app}(\theta)$ relation.

\subsection{Comparison of EGRET and non-EGRET Sources}
One of the major questions to be answered about the EGRET sources is
why some of the sources sharing the common characteristics of EGRET blazars
are not detected in $\gamma$-rays.  Some possible reasons for this mentioned
by von Montigny et al. (1995b) are that there may be intrinsic differences between the
detected and undetected sources; that all blazars emit $\gamma$-rays but some are not
currently seen due to long timescale variability; and that the $\gamma$-ray emission
may be beamed more narrowly than the radio emission.  VLBI observations are uniquely
suited to testing this third possibility.  Many of the source properties observed
with VLBI, including the apparent superluminal velocity, $\beta_{app}$, 
depend on the angle of the jet to the line-of-sight,
so in principle the jet orientation angles of the EGRET and non-EGRET sources can
be compared.  Note that the EGRET and non-EGRET sources do not necessarily have a
bimodal distribution in their ratio of $\gamma$-ray flux to radio flux, as the radio-loud
and radio-quiet quasars do in their ratio of radio to optical flux (e.g. Rawlings 1994).
However, since Mattox et al. (1997) show that there is a correlation between average
5 GHz radio flux and peak $\gamma$-ray flux, and many of the strongest radio blazars
have not been detected by EGRET, there is at least a very large scatter in the ratio of
$\gamma$-ray flux to radio flux that remains to be explained.  

We decided to compare the average values of $\beta_{app}$ for the EGRET and
non-EGRET sources to see if there is a significant difference between the two
groups.  We use the speeds of all VLBI components measured in this work, Paper I,
and Paper II, along with the collected apparent velocity data of VC94
and some more recent VLBI observations of the EGRET sources 0420-014 (Wagner et al. 1995),
0528+134 (Krichbaum et al. 1995; Pohl et al. 1995), and 1633+382 (Barthel et al. 1995).
Recent observations of 1730-130 (Bower et al. 1997) are not included since the
superluminal motion is only inferred from the timing of flux outbursts.
A similar comparison has been done by Tingay et al. (1996);
however, we include the speeds of 15 new VLBI components measured in this paper,
Paper I, and Paper II --- thus approximately doubling the available sample for EGRET sources ---
and we also use the data of VC94 somewhat
differently.  For the EGRET sources, we exclude the BL Lacs, and we consider only
the core-dominated quasars from VC94, because Gabuzda et al. (1994)
show that the $\beta_{app}$ values of BL Lacs and quasars are quite different.
We also combine the apparent velocity data using weighted averaging, so we
exclude any apparent velocity measurement published without an associated error.
When multiple components have been measured in a single source, we take the weighted
average of the component speeds to form an average value for that source; this is exactly
equivalent to forming the weighted average of {\em all} components from {\em all} sources.
Since VC94 remark that the large number of stationary
components found in core-dominated quasars implies that they probably represent
a different phenomenon from the more rapidly moving components, we also exclude 
components where the velocity measurement and associated error are consistent with
no motion.  We find that the two groups of EGRET and non-EGRET sources have the same
average apparent velocity of 5.5$h^{-1}c$, with the standard deviations of the
distributions being 2.7 and 2.4 $h^{-1}c$ respectively.  
Histograms of these apparent velocities are shown in Piner (1998).  We find no evidence that
the EGRET and non-EGRET sources have different apparent velocities, in agreement with
Tingay et al. (1996).
    
Apparent velocity is probably not the best indicator of orientation.
The graph of $\beta_{app}$ vs. $\theta$ has a maximum at $\sin\theta=
1/\Gamma$, and below this maximum the apparent velocity decreases as the
source moves closer to the line-of-sight.  If a significant number of
EGRET sources are located on the small angle side of this maximum, they
could actually have a lower average apparent velocity than other sources
while still being closer to the line-of-sight.  A better indicator of orientation
is a function which increases monotonically as the angle to the line-of-sight
decreases, such as the Doppler factor $\delta$ or the ratio of core to extended
flux, or core dominance parameter, $R$.  

We have investigated the average values of $\delta$ and $R$ for EGRET and
non-EGRET sources using the values of $\delta_{SSC}$ from GPCM, DG95,
GD96, Guerra \& Daly (1997), and this paper;
values of $R$ from GPCM, DG95, and Murphy et al. (1993);
and using the set of EGRET sources from Mukherjee et al. (1997).  The average values for
$\delta_{SSC}$ and $R$ of the EGRET and non-EGRET sources are given in Table~\ref{deltar}, where
we have used the same source classifications used by GPCM.  The average values of
$\delta_{SSC}$ for the EGRET and non-EGRET sources are the same within the errors,
which may simply reflect the large uncertainties in determining $\delta_{SSC}$.
The only significant difference in the $R$ values occurs for the LPQs, where the
EGRET LPQs have an $R$ value significantly higher than the non-EGRET LPQs.
This implies that the EGRET LPQs are more strongly beamed than the non-EGRET
LPQs and, if degree of polarization increases with decreasing angle to the line-of-sight,
implies that the EGRET LPQs have an angle to the line-of-sight typical of HPQs.
Since degree of polarization is variable, this suggests that the EGRET LPQs may 
really be HPQs which were classified during a low-polarization state.

\begin{table}[!h]
\caption{Average Values of $\delta_{SSC}$ and $\log R$}
\label{deltar}
\begin{tabular}{c c c c c} \tableline \tableline
& \multicolumn{2}{c}{EGRET} & \multicolumn{2}{c}{non-EGRET} \\
& \multicolumn{2}{c}{sources} & \multicolumn{2}{c}{sources} \\
Source Type & $\delta_{SSC}$\tablenotemark{b} & $\log R$ 
& $\delta_{SSC}$ & $\log R$ \\ \tableline
BL Lacs & 4.0$\pm$1.5 & 1.76$\pm$0.38 & 3.1$\pm$0.7 & 1.63$\pm$0.21 \\ 
CDQs\tablenotemark{a} & 7.9$\pm$1.4 & 1.27$\pm$0.15 & 7.3$\pm$1.0 & 1.08$\pm$0.10 \\ 
HPQs & 8.5$\pm$1.9 & 1.21$\pm$0.19 & 9.0$\pm$1.3 & 1.33$\pm$0.21 \\ 
LPQs & 7.1$\pm$2.1 & 1.43$\pm$0.25 & 6.8$\pm$1.7 & 0.93$\pm$0.10 \\ \tableline
\end{tabular}
\tablenotetext{a}{Core Dominated Quasars.  Combination of the HPQs and LPQs.}
\tablenotetext{b}{Errors given are the standard deviation of the mean.}
\end{table}

We also compared the misalignment angle distribution of the EGRET blazars to that
of the non-EGRET blazars, where the misalignment angle is the difference between
the VLBI and VLA-scale position angles.  For the EGRET sources we measured 
misalignment angles using the method of Xu et al. (1994) 
for all sources where images were available in the literature, 
as well as using the new images from this paper, Paper I, and Paper II.
A complete listing of these misalignment angles is given by Piner (1998).
We compared this misalignment angle distribution to that for HPQs in general
using the results from the combined Pearson \& Readhead and first Caltech-Jodrell Bank surveys
presented by Xu et al. (1994) and found that EGRET sources 
do not preferentially belong to either the aligned or
the misaligned population, but follow the distribution typical of highly polarized
sources in general.  These results independently confirm similar results found
recently by Bower et al. (1997).
 
In conclusion, we do not find strong evidence that we are viewing $\gamma$-ray sources
any closer to the line-of-sight than is typical for an HPQ source.  
Moellenbrock et al. (1996), in a survey of core-dominated radio sources, 
found that the $\gamma$-ray sources occupy the high
end of their brightness temperature distribution, indicating
that the $\gamma$-ray sources are among the most highly beamed in their sample.
However, their sample includes both high and low polarization core-dominated
quasars.  Unified schemes (Guerra \& Daly (1997), Ter\"{a}sranta \& Valtaoja (1994))
indicate that LPQs typically have a larger angle to the line-of-sight than HPQs.
If the KS test done by Moellenbrock et al. (1996) is redone using only
the HPQ sources from their sample and the latest list of EGRET sources, the significance
of the difference between the brightness temperature distributions of the EGRET and
non-EGRET sources drops below 95\%.
We conclude that VLBI observations are consistent with the average opening angle for the $\gamma$-ray
emission being approximately equal to the average viewing angle for an HPQ.
If this is so, then the reason some HPQs are not seen in $\gamma$-rays
is probably due to intrinsic source differences or time variability.  Nair (1997) has stated
that cluster analysis shows that $\gamma$-ray sources have larger amplitudes of optical
variability and are bluer at larger redshifts.  Punsly (1996) has found that $\gamma$-ray
quasars have larger mm spectral peaks relative to their cm spectral peaks than other
core-dominated quasars.  Both of these may be evidence that there are some intrinsic
differences between the EGRET and non-EGRET sources.

\section{Conclusions}
We have presented VLBI images of the EGRET quasars 0202+149, CTA 26, and 1606+106.
We have detected superluminal motion in two of these sources, CTA 26 and 1606+106;
such superluminal motion is expected from highly beamed $\gamma$-ray blazars.
The high Doppler factor derived for 0202+149 also indicates that it is strongly
beamed, the stationary jet components in this source can be interpreted as standing
shocks or flow very close to the line-of-sight.  The quasar 0202+149 satisfies
all the criteria of the compact F double morphology class as given by PR88.
We have also investigated the shapes of the VLBI jets in all of these sources,
as well as in 1156+295, and have found they all possess apparently bent jets.
We have detected non-radial motions of components in CTA 26 and 1156+295.
It has been noted by some authors (Krichbaum et al. 1995; Wehrle et al. 1996), and is
expected from some theoretical models, that VLBI components appear subsequent to
$\gamma$-ray flares.  We have not yet detected any 
components emerging subsequent to the $\gamma$-ray flares in
CTA 26, 1156+295, and 1606+106, although the only source which has a lower limit
on the ejection time of such a component that is significantly later than the $\gamma$-ray flare
in that source is 1156+295.

The observations of the three sources presented here, along with the observations
of 1611+343 and 1156+295 presented in Papers I and II, significantly increase the
number of EGRET sources for which detailed VLBI monitoring data are available.  
Such an increase makes comparisons between properties of the EGRET and non-EGRET populations
more significant.  We have compared the misalignment angle distribution of EGRET sources
to the distribution for blazars as a whole and find that the EGRET sources do
not preferentially belong to the aligned or the misaligned population.  
We have also compared the average values of the apparent velocity, the Doppler factor,
and the core dominance parameter for the EGRET sources with the same quantities for the
non-EGRET sources.  We find no significant difference in these quantities between the two
groups. We thus find no indication that the EGRET blazars are more strongly beamed 
than the non-EGRET blazars,
although the estimated errors and the scatter of individual sources are rather
large, particularly for estimates of the Doppler factor.
This paper, along with Paper I and Paper II, shows the great usefulness of archived geodetic
VLBI data in producing high quality, densely time sampled series of VLBI images for use
in astrophysical studies.  The Washington VLBI correlator's geodetic database continues to
be a potential source for such studies. 

\vspace{0.5in}
We acknowledge the support of the VLBI staff at the Naval Observatory
and the EGRET team at NASA/GSFC.
We also acknowledge helpful discussions and communication of data in
advance of publication from Margo Aller, Steve Bloom, Alan Fey, and Bob Hartman.
Additionally, we acknowledge the referee for his/her helpful comments and
suggestions, particularly in pointing out the dangers of mentally equating EGRET and
non-EGRET blazars to radio-loud and radio-quiet quasars.
This research has made use of the United States Naval Observatory (USNO) Radio
Reference Frame Image Database (RRFID),
data from the University of Michigan Radio Astronomy Observatory which is supported by
the National Science Foundation and by funds from the University of Michigan,
and the NASA/IPAC extragalactic database (NED)
which is operated by the Jet Propulsion Laboratory, Caltech, under contract
with the National Aeronautics and Space Administration.

\end{document}